\documentclass[fleqn,usenatbib]{mnras}
\usepackage{newtxtext,newtxmath}
\usepackage[T1]{fontenc}
\usepackage{ulem}

\DeclareRobustCommand{\VAN}[3]{#2}
\let\VANthebibliography\thebibliography
\def\thebibliography{\DeclareRobustCommand{\VAN}[3]{##3}\VANthebibliography}

\usepackage{graphicx}

\title[Collision rates of planetesimals]{Collision rates of planetesimals near mean-motion resonances}

\author[S. C. Wallace et al.]{
Spencer C. Wallace$^{1}$\thanks{E-mail: scw7@uw.edu (SCW)},
Thomas. R. Quinn$^{1}$,
Aaron C. Boley$^{2}$
\\
$^{1}$Astronomy Department, University of Washington, Seattle, WA 98195\\
$^{2}$Department of Physics and Astronomy, University of British Columbia, Vancouver BC, Canada
}

\date{Accepted XXX. Received YYY; in original form ZZZ}

\pubyear{2020}

\begin{document}
\label{firstpage}
\pagerange{\pageref{firstpage}--\pageref{lastpage}}
\maketitle

\begin{abstract}
In circumstellar discs, collisional grinding of planetesimals produces second-generation dust. While it remains unclear whether this ever becomes a major component of the total dust content, the presence of such dust, and potentially the substructure within, it can be used to explore a disc's physical conditions. A perturbing planet produces nonaxisymmetric structures and gaps in the dust, regardless of its origin. The dynamics of planetesimals, however, will be very different than that of small dust grains due to weaker gas interactions. Therefore, planetesimal collisions could create dusty disc structures that would not exist otherwise. In this work, we use N-body simulations to investigate the collision rate profile of planetesimals near mean-motion resonances. We find that a distinct bump or dip feature is produced in the collision profile, the presence of which depends on the libration width of the resonance and the separation between the peri- and apocenter distances of the edges of the resonance. The presence of one of these two features depends on the mass and eccentricity of the planet. Assuming that the radial dust emission traces the planetesimal collision profile, the presence of a bump or dip feature in the dust emission at the 2:1 mean-motion resonance can constrain the orbital properties of the perturbing planet. This assumption is valid, so long as radial drift does not play a significant role during the collisional cascade process. Under this assumption, these features in the dust emission should be marginally observable in nearby protoplanetary disks with ALMA.
\end{abstract}

\begin{keywords}
planet-disc interactions -- planets and satellites: formation -- submillimetre: planetary systems -- methods: numerical
\end{keywords}

\section{Introduction} \label{sec:intro}

Recent observations of circumstellar discs by ALMA have revealed a rich variety of substructure. Features such as gaps and asymmetries 
\citep{2015ApJ...808L...3A, 2016Sci...353.1519P, PhysRevLett.117.251101, 2016ApJ...820L..40A, 2016Natur.535..258C} in the emission provide 
diagnostics for the physical processes that drive the evolution of the discs. In many gas-rich systems, these gap features are argued to indicate the 
presence of a giant planet, either embedded in the disc \citep{2015MNRAS.453L..73D} or orbiting adjacent to it, although the details of gas-dust 
interactions can have strong effects on the resulting morphologies
\citep{2018ApJ...866..110D}. In gas-poor/debris systems, a giant
planet perturber can also influence 
the structure of the dust continuum emission. For example, a misaligned giant planet can produce nonaxisymmetric features such as warps 
\citep{2001A&A...370..447A}, highly eccentric perturbers can produce structures through secular interactions 
\citep{2014MNRAS.443.2541P, 2015MNRAS.448.3679P}, and mean motion resonances (MMRs) can open gaps
\citep{2015ApJ...798...83N, 2016ApJ...818..159T, 2018ApJ...857....3T}.

Collisions between planetesimals are thought to be the principal source of dust for debris discs \citep[see][]{2008ARA&A..46..339W}.  Although some 
amount of primordial dust is likely still present during the early stages of debris disc evolution, ongoing collisions between small bodies will augment 
this and could be used to trace the underlying dynamical activity of the planetesimals. This process was directly detected when the putative object 
Fomalhaut b was confirmed to be an expanding debris cloud, most likely from a planetesimal collision \citep{2020PNAS..117.9712G}. 

\citet{2013ApJ...777L..31D} showed that collisional dust that is generated near the gap opened by a giant planet in the inner regions of a transition disc should 
produce a distinct observable marker that could be used to infer the presence of the planet. In a followup study, it was found that morphological 
differences in the dust emission could be used to determine whether the planet had a circular or eccentric orbit \citep{2016ApJ...820...29D}. Moreover, 
substructure due to mean-motion resonances (MMRs) with a giant planet may hold additional clues that could be used to constrain the properties of 
the planet.  The width of a resonance is set by both the mass of the perturbing planet and the unperturbed eccentricity of the planetesimals, 
which is in turn set by the eccentricity of the planet through secular forcing. Therefore, the planetesimal collision profile and potentially the structure of second generation 
dust produced near MMRs may encode information about both the mass and eccentricity of the planet.

The dynamics governing the motion of bodies near MMRs is extremely nonlinear, as is determining what the collision rates between planetesimals 
should look like in these regions. For a collection of bodies massive enough to experience the effects of gravitational focusing, a large eccentricity 
dispersion tends to reduce the probability of collision, while enhancements in surface density tends to increase it. Due to conservation of the Jacobi 
energy, MMRs simultaneously enhance the local eccentricity dispersion and also enhance the surface density adjacent to the resonance 
\citep{2000Icar..143...45R, 2017ApJ...850..103B}. Unfortunately, collision detection in an N-body simulation is extremely computationally expensive. 
So far, studies of planetesimal dynamics near MMRs in which the planetesimals are directly resolved have involved either collisionless test particles 
\citep{2017ApJ...850..103B, 2016ApJ...818..159T, 2018ApJ...857....3T} or integration times that do not fully capture the dynamics of the resonances 
\citep{2000Icar..143...45R, 2013ApJ...777L..31D}.

To further elucidate this subject, we use the tree based N-body code {\sc ChaNGa} \citep{2008IEEEpds...ChaNGa, 2015AphCom..2..1} to follow the 
collisional evolution of a planetesimal disc under the gravitational influence of a Jupiter-sized body. Because particle positions are sorted into a tree 
structure, neighbor finding and collision detection can be done quickly and efficiently. This considerably relaxes the constraints on resolution and 
integration time. With this toolset, we explore the collision rate structure of a planetesimal disc in the vicinity of mean-motion resonances with a planet. In 
particular, we would like to determine (1) what dynamics govern the collision rate profile near MMRs, (2) whether MMRs leave a detectable signature in the
collisionally-generated dust, and (3) whether these signatures can be used to determine or constrain the orbital properties of the perturbing planet.

This work is organized in the following way: In section \ref{sec:dynamics}, we provide an overview of the relevant dynamics that drive the evolution 
of a planetesimal disc under the gravitational influence of an external perturber. In section \ref{sec:sims} we provide an overview of the N-body code 
used and describe the initial conditions chosen for five simulations in which a perturbing giant planet is given various masses  and eccentricities. Section
\ref{sec:results} presents the results of these simulations and we take an in-depth look at the collision rate profiles of the planetesimals near the MMRs. 
In section \ref{sec:dust},  we extrapolate the collision rate down to smaller bodies and motivate a direct correspondence between the planetesimal collision profile
and the resulting dust profile. In section \ref{sec:constrain}, we use the resolved collisions to generate synthetic dust 
emission profiles that would be detected with observing facilities like ALMA. Under our simplifying assumptions, we show that a characteristic bump 
or dip feature appears in the dust emission near the interior 2:1 MMR, the presence of which depends on the mass and eccentricity of the perturbing 
planet. If the 2:1 MMR can be identified (presumably, by identifying another prominent MMR, such as the 3:1, and measuring the spacing between the two), we discuss the 
potential for using this feature to place constraints on the mass and eccentricity of the planet. In addition, we highlight the caveats of connecting the dust and planetesimal collision 
profiles in such a simple way. Finally, we conclude in section \ref{sec:conclusions}.

\section{Overview of Relevant Dynamics} \label{sec:dynamics}

We begin by providing a description of the dynamical effects responsible for shaping the orbital distribution of the planetesimals. The purpose of this is twofold: (1) to motivate
the initial conditions used for the simulations described in section \ref{sec:ics} and (2) to justify the exclusion of certain physical effects from our simulations. Here, we focus solely on the 
physics relevant for full-sized ($\sim$100 km) planeteismals and save a discussion of the effects on smaller bodies generated through collisions for section \ref{sec:dust}.

\subsection{Secular Forcing}\label{sec:sec_force}

The most direct and widespread effect that a perturbing giant planet will have on a planetesimal disc is through secular forcing of the planetesimals. This will 
cause the complex eccentricities of the planetesimals to take on a time-independent 
forced value, given by \citep{1999ApJ...527..918W} as

\begin{equation}\label{eq:eforced}
	z_{f} = \frac{b^{2}_{3/2} (\alpha)}{b^{1}_{3/2} (\alpha)} e_{g} ~ \mathrm{exp} ~ i \varpi_{g}.
\end{equation}

\noindent Here, $\alpha = a_{g} / a$ where $a_{g}$ and $a$ are the semi-major axes of the giant planet and a planetesimal, 
respectively. $e_{g}$ and $\varpi_{g}$ are the eccentricity and
longitude of pericenter of the giant planet, and $b^{j}_{s} (\alpha)$ is a 
Laplace coefficient given by \citep{1999ssd..book.....M} (ch. 6, pg. 237, eq. 6.67) as

\begin{equation}\label{eq:lap}
	b_{s}^{j}(\alpha) = \frac{1}{\pi} \int_{0}^{2 \pi} \frac{\cos \, j \psi \, d \psi}{\left( 1 - 2 \alpha \, \cos \psi + \alpha^2 \right)^{s}}.
\end{equation}

Without any nearby secular or mean motion-resonances, equation \ref{eq:eforced} will completely describe the eccentricities and longitude of  
pericenter orientations of the planetesimals. Additional forces due to two-body scattering between planetesimals, along with 
aerodynamic gas drag will add an additional free component to the complex eccentricity, which will be randomly oriented. The 
magnitude of the free eccentricity describes how dynamically hot the planetesimal disc is and sets the random encounter speeds of 
planetesimals. When the dynamical excitation of the disc is driven by gravitational stirring, the magnitude of the free eccentricity can 
be described by a Rayleigh distribution \citep{1992Icar...96..107I}.

For the case of a Jupiter mass planet at 5.2 au perturbing a test particle at 3 au, the timescale for secular forcing is approximately 12,000 years. The details of this calculation
can be found in appendix \ref{sec:sec_forcing_timescale}.

\subsection{Mean-Motion Resonances}\label{sec:mmr}

In regions where there are commensurabilities between frequencies, Laplace-Langrange secular theory breaks down and bodies are subject to strong 
perturbations. For the purposes of this study, we will ignore secular resonances, which generally occur on rather large timescales and will focus on 
mean-motion resonances. A MMR occurs  when the orbital period ratio between two bodies is sufficiently close to

\begin{equation}\label{eq:per_mmr}
	\frac{P}{P'} = \frac{p + q}{p},
\end{equation}

\noindent where  $p$ and $q$ are integers $>$ 0 and the unprimed and primed quantities correspond to the perturber and the body being perturbed, 
respectively. In terms of these quantities, $P/P'$ corresponds to a $p+q:p$ resonance. If the perturber is much more massive than the other body and all of the 
bodies lie in a near-Keplerian potential, the condition for MMR is set by

\begin{equation}\label{eq:a_mmr}
	\frac{a}{a'} = \left( \frac{p}{p + q} \right)^{2/3}.
\end{equation}

If we further assume that the two bodies are orbiting in the same plane, the motion of the bodies near resonance is determined by the
behavior of a critical angle

\begin{equation}\label{eq:phi_crit}
	\phi = (p + q) \lambda' - p \lambda - q \varpi,
\end{equation}

\noindent where $\lambda = \varpi + M$ is the mean longitude of a
body, with $M$ being the mean anomaly. For bodies in resonance, the critical angle will librate around an 
equlibrium value, while this angle will circulate outside of resonance. For small eccentricities, this behavior is analogous to the motion of a 
pendulum. Furthermore, variations in the critical angle are coupled to changes in the mean motion and semimajor axis \citep{1999ssd..book.....M}. An important 
point to note, which we will revisit later, is that the variation frequency of this angle approaches zero near the edge of a resonance. The 
width of a resonance can be defined by determining the largest variation in semimajor axis that permits librational, rather than circulatory motion of $\phi$. 
Calculations for the widths of first and second order interior MMRs are shown in appendix \ref{sec:libration}, along with a timescale for the libration. For the most prominent interior 
mean-motion resonances, including the 2:1 and 3:1, the timescale associated with these oscillations driven by a Jupiter mass planet is  $\sim$ 1,000 - 2,000 years.




\subsection{Collisions Between Planetesimals}\label{sec:colleq}

A simple analytic model for the collision rate of a planetesimal population is given by \citet{1969edo..book.....S} as

\begin{equation}\label{eq:saf}
	n \sigma v = n \pi s^{2} \left( 1 + 2 G m / s v^{2} \right) v,
\end{equation}

\noindent where $n$ is the number density of the population, $s$ and $m$ are the radii and masses of the bodies and $v$ is their typical encounter velocity. The encounter velocity is often described in terms of the rms eccentricity $\left<e^{2}\right>^{1/2}$ and inclination $\left<i^{2}\right>^{1/2}$ of the population (e.g. \citet{1993ARA&A..31..129L}) as

\begin{equation}\label{eq:eccincvel}
	v = \sqrt{\left< e^{2} \right> + \left< i^{2} \right>} v_{k},
\end{equation}

\noindent where $v_{k}$ is the local Keplerian velocity. The second term in equation \ref{eq:saf} can be thought of as an additional enhancement to the collision cross section due to 
gravitational focusing. When the typical encounter velocity is small compared to the mutual escape velocity of the planetesimals, the collision cross section greatly exceeds the geometric 
value. An important feature of equation \ref{eq:saf} is that for a fixed value of $n$, the collision rate exhibits a global minimum as a function of $v$. For small $v$, gravitational focusing 
facilitates more collisions, while for large $v$ the encounter rate simply becomes so great that the collision rate again increases, even though gravitational focusing is mostly suppressed. 
This point will become relevant in section \ref{sec:vary_ecc} when we examine the qualitative changes in the collision rate near mean-motion resonances.

Although equation \ref{eq:saf} works well to describe the collision rate for a homogeneous collection of planetesimals, some problems arise when regions of commensurabilities are 
introduced. Namely, the resonances cause the orbits of planetesimals to precess, and an interface between secularly aligned and randomly oriented orbits arise on each side of the 
resonance. Additionally, an interface between dynamically cold and hot planetesimals develop. These two effects cause the number density, collision cross section and encounter velocity 
to rather abruptly vary across the boundaries of the resonance. For this reason, an N-body treatment in which collisions are directly resolved is necessary to understand how the collision 
rate varies near the MMRs.

\subsection{Gas Drag on Planetesimals}\label{sec:pl_drag}

Over the course of many orbits, the residual gas from the primordial nebula can damp the eccentricities and inclinations of bodies. For a planetesimal-sized body, gas drag operates 
in the Stokes regime and the timescale for aerodynamic forces to significantly alter its relative velocity is given by \citep{1976PThPh..56.1756A}

\begin{equation}\label{eq:ts_stokes}
    t_{s} = \frac{2 m}{C_{D} \pi s^{2} \rho_{g} v_{g}},
\end{equation}

\noindent where $m$ and $s$ are the mass and radius of the planetesimal. $C_{D}$ is a drag coefficient which is of order unity, $
\rho_{g}$ is the local density of the gas and $v_{g}$ is the headwind velocity of the gas experienced by the planetesimal. At 3 au, the gaseous component of the solar nebula has a
density of $3 \times 10^{-11}$ g cm$^{-3}$ and the typical headwind experienced by a body on a Keplerian orbit is $\sim 5,000$ cm s$^{-1}$ \citep{1981PThPS..70...35H}. For a 100 km 
body with a density of 2 g cm$^{-3}$ (assuming a mixture of ice and rock) the stopping timescale is about 1 Myr. This is much longer than the timescales associated with 
secular forcing and libration due to mean-motion resonances, as discussed above, but shorter than the typical protoplanetary disk lifetime and potentially shorter the planetesimal 
formation timescale. For this reason, we do not model the effects of gas drag in the simulations, although we consider its effects when constructing initial conditions, which is discussed in 
the next section.

\begin{figure*}
\begin{center}
    \includegraphics[width=\textwidth]{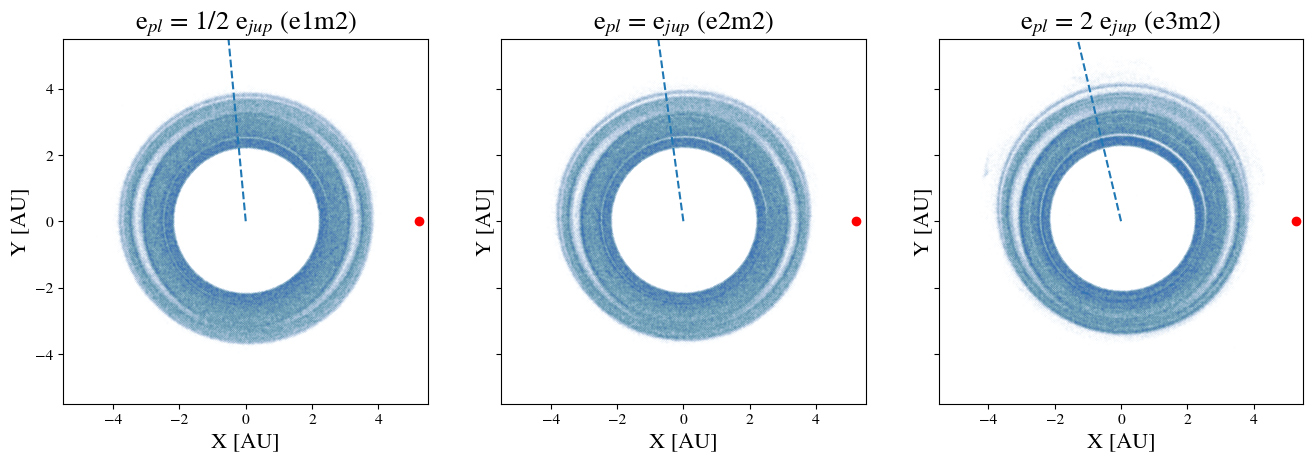}
    \caption{The positions of the remaining planetesimals at the end of the e1m2 (left), e2m2 (center) and e3m2 (right) simulations. The red dot 
    indicates the position of the giant planet, and the dashed line points in the direction of the planet's longitude of perihelion. Non-axisymmetric gaps 
    are apparent near the locations of MMRs. At higher eccentricities, more resonances become visible. At the 2:1 MMR, gap features at $\theta$ = 0 
    and $\theta = \pi$ follow the giant planet in its orbit.\label{fig:xy}}
\end{center}
\end{figure*}

\section{Simulations} \label{sec:sims}

\subsection{Numerical Methods}\label{sec:methods}

To follow the dynamical and collisional evolution of a planetesimal disc, we use the highly parallel N-body code {\sc ChaNGa} 
\footnote{A public version of {\sc ChaNGa} can be downloaded from \url{http://www-hpcc.astro.washington.edu/tools/ChaNGa.html}}. 
This code, which is written in the {\sc CHARM++} parallel programming language, was originally designed for cosmology simulations 
and has been shown to perform well on up to half a million processors \citep{2015AphCom..2..1}. Using similar methods to {\sc PKDGRAV}, which has been used for numerous 
studies of planet formation \citep{2000Icar..143...45R, 2005ApJ...625..427L, 2013ApJ...777L..31D, 2015ApJ...806...23L}, {\sc ChaNGa} calculates 
gravitational forces using a modified Barnes-Hut \citep{1986Natur.324..446B} tree with hexadecapole expansions of the moments and integrates the equations of motion using a 
kick-drift-kick leapfrog scheme. All of the simulations we perform use a node opening criterion of $\theta_{BH}$ = 0.7. More information about the code can be found in 
\citet{2008IEEEpds...ChaNGa}.

We have recently modified {\sc ChaNGa} to handle solid-body collisions between particles, by assigning them a fixed radius, rather than treating them as tracers of a fluid with a 
characteristic softening length. We provide a brief summary of the collision model here, although a full description is provided in \citet{2019MNRAS.489.2159W}. This work is largely based 
on the solid-body collision implementation in {\sc PKDGRAV}, which is detailed in \citet{1994MNRAS.269..493R} and \citet{2000Icar..143...45R}. Imminent collisions are found during the 
drift phase of each time step by extrapolating the positions of bodies forward using the velocity calculated during the first kick. For each body, the nearest 64 neighbors are tested for an 
imminent collision. If an imminent collision is detected, the two particles of mass $m_{1}$ and $m_{2}$ are merged together together to form one single larger body with the same density 
and a mass of $m_{1} + m_{2}$. The resulting body is then imparted
with the center of mass position and velocity of the two
colliders. Because the resolution of a collision can result in 
another imminent collision, these events must be handled one at a time, with the soonest collision being resolved first. For this reason, another collision search is run each time a collision 
is resolved, which continues until there are no more imminent collisions during the current time step. During the course of the simulation, no debris is created by collisions. To model the 
collisionally generated dust distribution, we use the statistics of
the resolved planetesimal collisions to build a dust profile. This process is discussed in detail in section \ref{sec:dust}.

\begin{table}
\begin{center}
\caption{Summary of Simulations Run}
\begin{tabular}{lllll} \hline \hline
Name     & Mass of Planet & Eccentricity of Planet &  &  \\ \hline
e1m2 & 1.0 $M_{jup}$                     & 0.5 $e_{jup}$                            &  &  \\
e2m2      & 1.0 $M_{jup}$                     & 1.0 $e_{jup}$                             &  &  \\
e3m2 & 1.0 $M_{jup}$                     & 2.0 $e_{jup}$                             &  &  \\
e2m1 & 0.5 $M_{jup}$                   & 1.0 $e_{jup}$                             &  &  \\
e2m3 & 2.0 $M_{jup}$                     & 1.0 $e_{jup}$                             &  &  \\ \hline
\end{tabular}
\label{tab:sims}
\end{center}
\end{table}

\subsection{Initial Conditions}\label{sec:ics}

In total, five simulations are run, which are listed in table \ref{tab:sims}. The second is a ``nominal'' case, in which the perturbing planet's mass and eccentricity are set to that of Jupiter's. 
In the other four cases, the mass or eccentricity is altered by a factor of two from the nominal value. In all cases, the perturbing giant is placed on a 5.2 au orbit around a 1 
$M_{\odot}$ star. The planetesimal disc extends from 2.2 to 3.8 au, which covers the two most prominent mean-motion resonances with the giant planet, the 2:1 at 3.27 au and the 3:1 at 
2.5 au. The planetesimal disc loosely follows a minimum-mass solar nebula surface density profile \citep{1981PThPS..70...35H}

\begin{equation}\label{eq:surf_den}
	\Sigma = \Sigma_{0} r^{-\alpha},
\end{equation}

\noindent with $\alpha$ = 3/2 and $\Sigma_{0}$ = 10 g cm$^{-2}$. Planetesimals are given a bulk density of 2 g cm$^{-3}$, which corresponds to a mixture of ice and rock, and 
are given a diameter of 300 km. This choice of parameters is meant to mimic the setup used in simulation B from \citet{2000Icar..143...45R}. One difference from the 
aformentioned study is that we use a narrower annulus, which reduces the number of particles required from $10^6$ down to roughly 500,000. This allows us to use a finer base timestep 
size, which, as we will discuss in section \ref{sec:results}, appears to be the reason why no features were seen in the collision profile near the resonances by \citet{2000Icar..143...45R}.

Another difference from \citet{2000Icar..143...45R} is that the dynamical effects of secular forcing by the giant planet, along with the effects of viscous stirring and gas drag on the 
planetesimals, are built into the initial conditions. Although we do not model the effects of gas drag on the planetesimals during the simulation, we construct the initial conditions 
such that the effects of gas drag and viscous stirring are in balance. The viscous stirring timescale of the planetesimal disc is much longer than our chosen integration time, and so the 
dynamical excitation of the disc (excluding resonances) stays constant, even without the inclusion of damping forces from the gas. This is done by first calculating the equilibrium 
eccentricity $e_{eq}$ due to viscous stirring and gas drag as a function of semimajor axis according to equation 12 of \citet{2002ApJ...581..666K}. The eccentricities of the bodies are 
drawn randomly from a Rayleigh distribution with a mode of $e_{eq}$, while the inclinations are drawn from a similar distribution with a mode of $e_{eq}$/2 \citep{1993MNRAS.263..875I}. 
The arguments of perihelion $\omega$, longitude of ascending nodes $\Omega$ and the mean anomalies $M$ of the bodies are drawn uniformly $\in [0, 2 \pi)$.

To account for the effects of secular forcing by the planet, the eccentricity vectors of the planetesimals are first decomposed into real and imaginary 
components:

\begin{equation}\label{eq:kh}
	z = (k, ih) = e \, exp(i \varpi)
\end{equation}

\noindent and a forced component is added to $h$ according to equation \ref{eq:eforced} (where we have set $\varpi_{g}$ = 0).

\subsection{Time Stepping Scheme}\label{sec:timestep}

For the purposes of the integrator, there are two relevant timescales in this system. The first is the orbital dynamical time $\sqrt{a^3/
G M_{\odot}}$. Through experimentation, we have found that using a base timestep size with {\sc ChaNGa} of 3\% of an orbital dynamical time
at the inner edge of the disc keeps the integration symplectic. Although doing so preserves orbital frequencies, the errors associated with
precession frequencies do not average out to zero. This is especially important given the effects of secular forcing by the planet. To mitigate
this, we further reduce the base timestep size by a factor of 4 and use $\Delta t$ = 0.0025 yr. Doing so prevents the longitude of perihelia of
planetesimals at the inner edge of the disc, where this artificial precession is most severe, from drifting by more than the intrinsic spread in $\varpi$ due to the
free eccentricity over the course of our integrations.

An additional timescale is set by the dynamical time of the planetesimals ($\sim 1/\sqrt{G \rho}$), which is about 45 minutes. This timescale must be resolved in order to properly 
follow close gravitational encounters between these bodies. To resolve the base time step and the dynamical timescale of planetesimals simultaneously, we use a two-tiered time 
stepping scheme, following\citep{2015ApJ...806...23L}. To start, all bodies are placed on the base time step. A first pass of collision detection is then run in which the radii of all bodies are 
inflated by a factor of 2.5. Any bodies with imminent collisions predicted using the inflated radii are placed on a time step that is a factor of 16 smaller than the orbital time step. Although 
this is still a factor of roughly 7 larger than the dynamical time of a planetesimal, we found no difference in the collision rate when using any smaller of a minimum time step size.The 
purpose of the two-tiered scheme is to properly resolve the gravitational interactions between any bodies that undergo a close encounter. This prevents the coarser base time step from 
reducing the effectiveness of gravitational focusing, while minimizing the additional computational expense.

\begin{figure}
\begin{center}
    \includegraphics[width=0.5\textwidth]{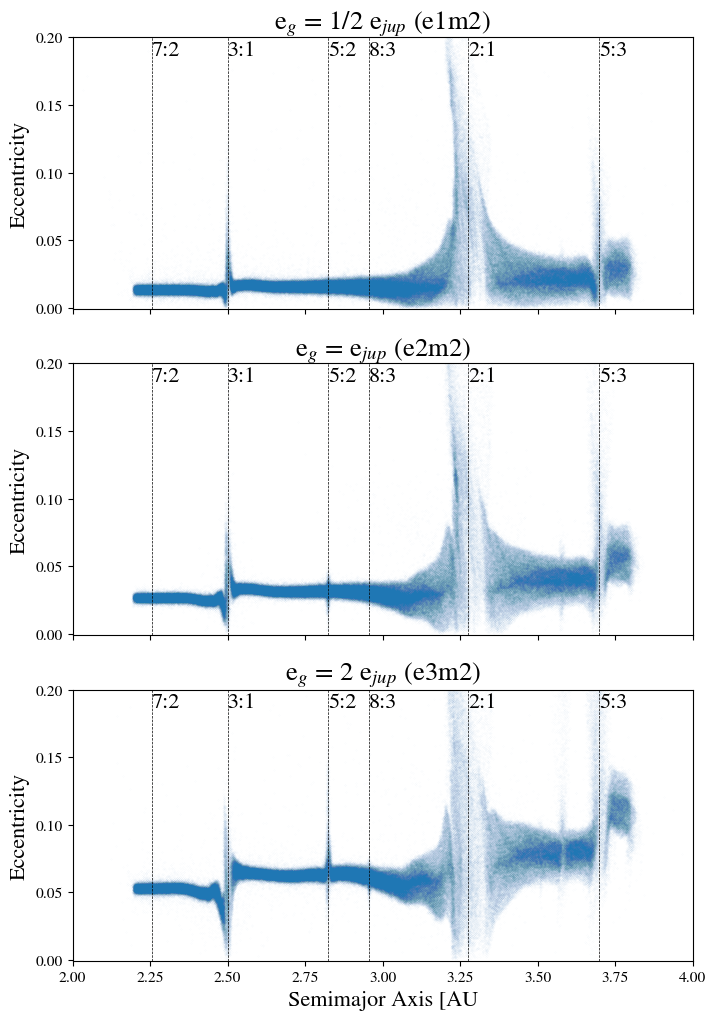}
    \caption{The semimajor axes and eccentricities of the remaining planetesimals are shown, with the locations of prominent resonances indicated 
    by the vertical dashed lines. Libration of the critical angle drives large variations in eccentricity, which produce spikes in the a-e plane. Between 
    the resonances, the nonzero eccentricity is due to secular forcing by the planet.\label{fig:ae}}
\end{center}
\end{figure}

\begin{figure*}
\begin{center}
    \includegraphics[width=\textwidth]{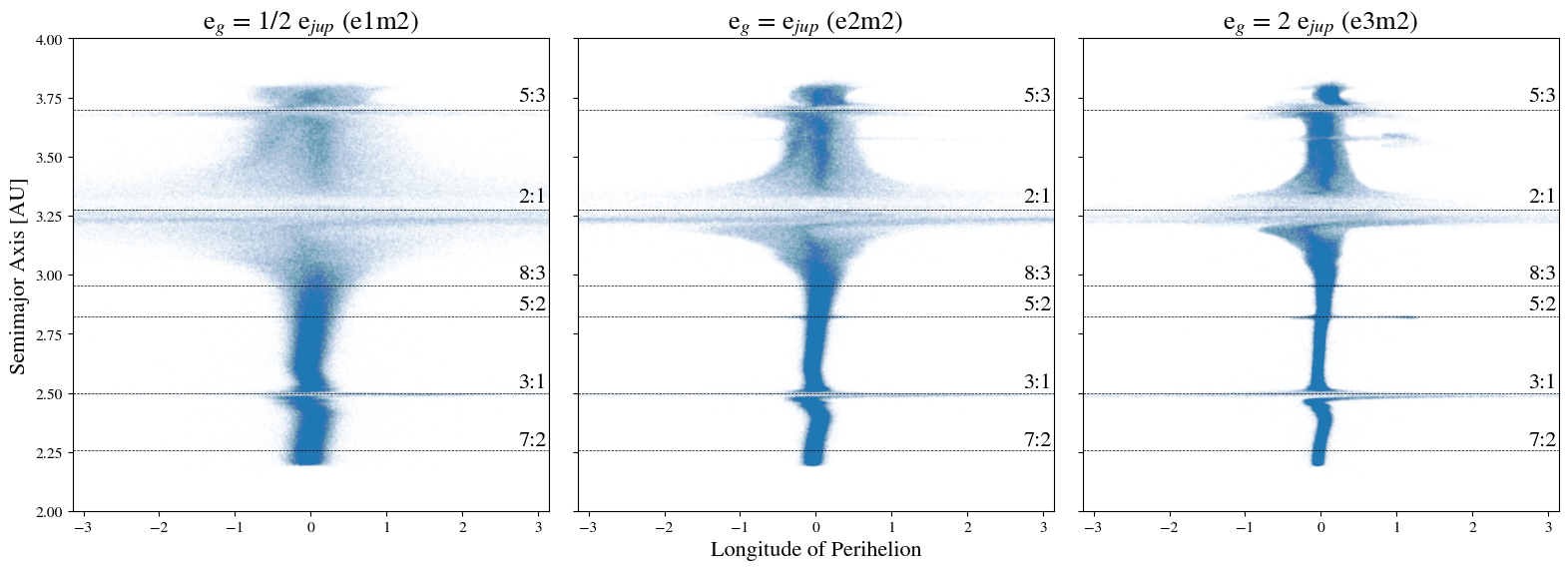}
    \caption{Shown here are the longitudes of pericenter and semimajor axes of the remaining planetesimals. The dashed lines again indicate the 
    locations of prominent MMRs. Close to the resonances, the critical angle librates and drives fast precession of the longitudes of pericenter of the 
    bodies. This overpowers the secular forcing by the planet and effectively randomizes the orientations of their orbits.\label{fig:long_ph}}
\end{center}
\end{figure*}

\section{Results} \label{sec:results}

All five simulations are evolved for 5,000 years, which is roughly 400 complete orbits of the perturbing planet. Because the effects of the mean 
motion resonances are not built into the initial conditions, the simulations must be run long enough for the distribution of orbital elements to reach 
equilibrium near the resonances before the collision rate is measured. To do so, we allow each simulation to run for 2,000 years before we begin recording any collision statistics. 
This is comparable to the libration period, which was calculated in section \ref{sec:mmr} for the 3:1 and 2:1 resonance. Although one might worry that a single libration period may not be 
long enough for the orbital structure due to the resonances to develop, we find that the shape of the semimajor axis-eccentricity distribution of the planetesimals reaches a steady state by 
this time. This is largely due to the fact that there are enough bodies in or near the resonances to allow an ensemble average of phase space, rather than a time average. As an additional 
confirmation, we find that the time evolution of the collision rate near the resonances maintains a steady state in all five simulations after $\sim$ 2000 years.

\subsection{Varying the Eccentricity} \label{sec:vary_ecc}

We begin by examining simulations e1m2, e2m2 and e3m2. The positions of the planetesimals in the x-y plane after 5,000 years of integration are 
shown in figure \ref{fig:xy}. In all cases, the coordinate system is rotated so that the giant planet lies at $\theta = 0$. The longitude of perihelion of 
the planet is shown by the dashed line. Resonances with the perturbing planet are visible as nonaxisymmetric gaps in this figure. Upon close inspection of a 
series of simulation snapshots, these gaps appear to follow the planet in its orbit, rather than aligning themselves with the longitude of perihelion. A 
similar substructure, which is most noticeable near the 2:1 MMR, reveals itself in \citet{2000Icar..143...45R} (see their figures 3c and 3f) and 
\citet{2016ApJ...818..159T} (see their figure 3a). This structure is also present in the \citet{2017ApJ...850..103B} simulations, although it was not reported at the 
time (see Appendix \ref{sec:boley_plot}). It is worth noting that both \citet{2000Icar..143...45R} and \citet{2017ApJ...850..103B} started with a completely cold planetesimal disc. 
Thus, the presence of these features seems robust to the choice of initial conditions.

The effects of the resonances become much more apparent in semimajor axis-eccentricity space, which is shown in figure \ref{fig:ae}. In all cases, 
the 3:1, 2:1 and 5:3 resonances are readily visible as ``spikes'' in the eccentricity that bend slightly inward (due to the conservation of the Jacobi energy). In the 
e3m2 simulation, features also appear near the 5:2, 7:3 and 5:3 resonances. The absence of these finer features from the e1m2 simulation can be 
explained by the fact that the strength of a resonance scales with $e^{q}$ \citep{1994PhyD...77..289M}.

Another important effect of the resonances is visible in figure \ref{fig:long_ph}, which shows the orientation of the longitude of perihelia of 
planetesimals in the disc. Inside of the resonances, orbits of planetesimals quickly precess, and their orientations are effectively randomized. An 
important point to note is that this strong precession effect quickly disappears beyond the boundaries of the resonance. This turns out to be 
key to explaining the nonaxisymmetric structure seen in figure \ref{fig:xy}, which will be addressed in more detail below.

\begin{figure}
\begin{center}
    \includegraphics[width=0.5\textwidth]{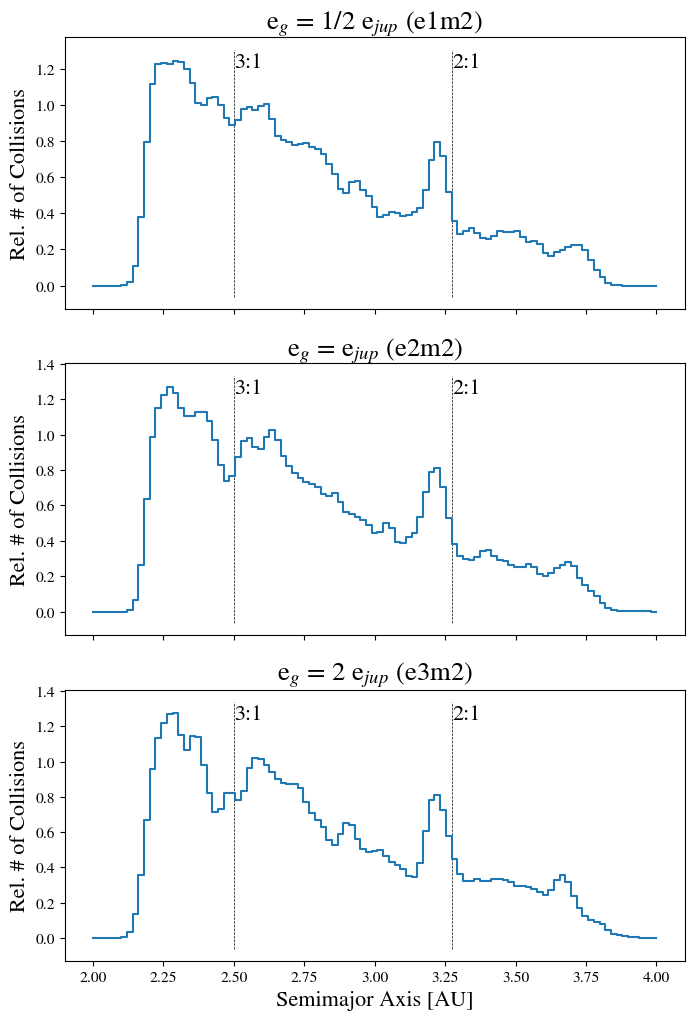}
    \caption{A PDF of the collision rate in each disc as a function of semimajor axis, generated using a KDE with a Gaussian kernel with a FWHM of 0.02 au. In semimajor axis space, 
    prominent features appear near the 3:1 and 2:1 MMRs. Near the 3:1, the collision rate exhibits a local minimum, while an enhancement appears near the 
    2:1.\label{fig:coll_hist_a}}
\end{center}
\end{figure}

Next, we examine the statistics of collisions resolved in each of the simulations. The 3D positions and velocities of the two colliding bodies are 
recorded to a table at the moment of impact. We derive the Keplerian orbital elements of a collision using these positions and velocities. 
First, we examine the semimajor axis of the first collider, which is shown in figure \ref{fig:coll_hist_a}. During a collision, the ``first'' particle is defined 
as the more massive of the two. However, nearly all of the collisions happen between the initial, equal-mass planetesimals. In this case, the 
distinction is set by the collision search algorithm and is rather arbitrary. In all of the plots where we show collision statistics, we have verified that 
using the ``first'' or ``second'' collider does not qualitatively change any of the features. We find that some of the features present in figure 
\ref{fig:coll_hist_a} in this and subsequent figures are highly sensitive to the number of bins and the location of the bin edges. For this reason, we 
construct a probability density function (PDF) of the collisions using a Kernel Density Estimate (KDE). We use the {\sc neighbors.KernelDensity} function from the {\sc sklearn} 
\citep{scikit-learn} package to construct our KDEs, using a gaussian kernel with a FWHM of 0.02 au. The curves shown in figure \ref{fig:coll_hist_a} are normalized such that the 
area underneath is equal to 1.

Near the stronger resonances, there are noticeable suppressions or enhancements to the local collision rate. This contrasts with the findings of 
\citet{2000Icar..143...45R}, who simulated a similar setup and found no discernible features near the MMRs. We attribute the differences to a more 
conservative timestepping criterion in our simulations, along with the inclusion of secular perturbations in our initial conditions. The most prominent features appear as a 
local maximum in the collision rate near the 2:1 MMR and a local minimum near the 3:1 MMR. At higher forced eccentricities, features near the 5:2, 8:3 and 5:3 
resonances are also visible, due to the steep sensitivity of higher order resonant perturbations to eccentricity \citep{1994PhyD...77..289M}.

Although the features in figure \ref{fig:coll_hist_a} near the 3:1 and 2:1 MMRs appear qualitatively different, they can be explained by one single dynamical process. As discussed in 
section \ref{sec:colleq}, the collision rate between planetesimals depends on both the encounter rate and the strength of gravitational focusing. The encounter rate grows as the relative 
velocity $v$ between the planetesimals increases, while gravitational focusing is most effective for bodies with small relative velocities. There is therefore an intermediate value of $v$ for 
which the average collision rate $\left< \sigma v \right>$ is at a minimum, which we will refer to as  $\left< \sigma v \right>_{0}$. The dynamical excitation induced by a mean motion 
resonance can either increase or decrease the local average collision rate, depending on the unperturbed value of  $\left< \sigma v \right>$ relative to  $\left< \sigma v \right>_{0}$.

\begin{figure*}
\begin{center}
    \includegraphics[width=\textwidth]{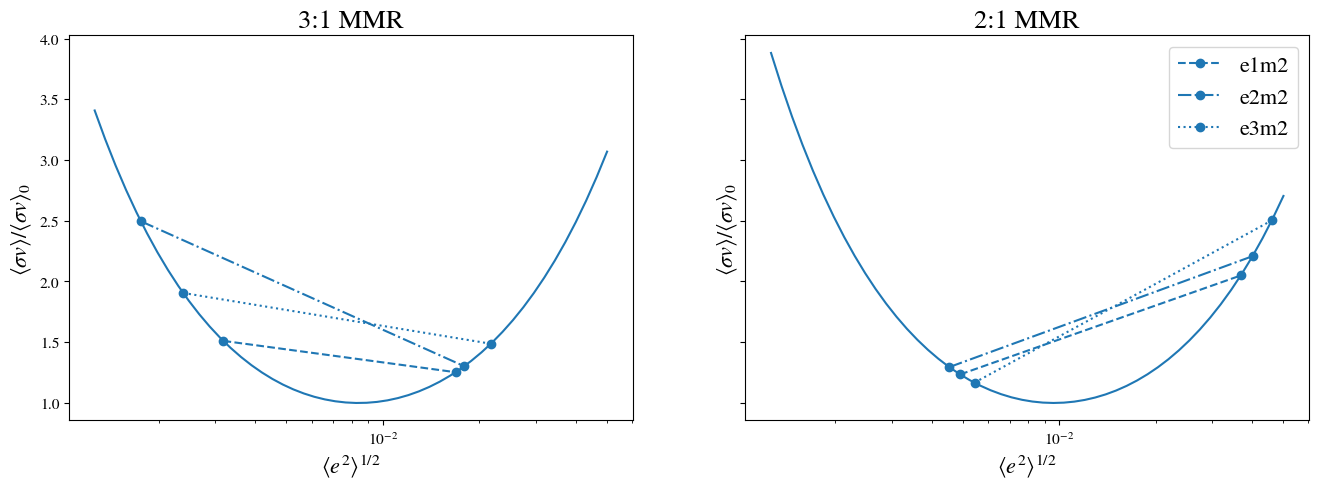}
    \caption{The collision rate (given by equation \ref{eq:saf}) of bodies in the vicinity of the 3:1 (left) and 2:1 (right) MMRs, relative to the minimum value, as a function of the local 
    eccentricity dispersion. The pairs of points connected by lines show the values of the unperturbed (left-hand points in each subplot) and perturbed (right-hand points in each subplot) collision rates in 
    the e1m2, e2m2 and e3m2 simulations. The unperturbed collision rate, along with the amount of dynamical heating that the planetesimal population experiences determines whether 
    the MMR acts to suppress or enhance the collision rate.\label{fig:gf}}
\end{center}
\end{figure*}

In figure \ref{fig:gf}, we show the effect that the 2:1 MMR and 3:1 MMR have on the local average collision rate. Each pair of points connected by a line represents the average 
collision rate before and after the effects of the mean-motion resonances develop. Here, the relative velocity between bodies is calculated by measuring the eccentricity and inclination 
dispersion near each resonance and using equation \ref{eq:eccincvel}. We assume that the eccentricity and inclination dispersions are coupled \citep{1993MNRAS.263..875I} and use the 
eccentricity dispersion as a free parameter. Because both the strength of secular forcing and the effects of the resonant perturbations are different near the 3:1 and 2:1 MMRs, the 
planetesimal population at each of these locations in the disk experiences a qualitatively different change in the collision rate. Near the 3:1 MMR, the unperturbed collision rate is well 
above the minimum value $\left< \sigma v \right>_{0}$ and the dynamical heating introduced by the resonance acts to suppress it further. Near the 2:1 MMR, the collision rate starts out 
much closer to the minimum value and the additional perturbations instead act to drive it higher.

This explains the relative drop in the collision rate near the 3:1 MMR and the relative increase near the 2:1 and 5:3 resonances seen in figure \ref{fig:coll_hist_a}. Although 
this could potentially serve as a useful diagnostic of the planetesimal size in a disc (because the size and mass of the planetesimals sets the eccentricity dispersion at which 
$\left< \sigma v \right>$ is minimized), a direct measurement of the semimajor axes of the colliding planetesimals is not possible. Furthermore, as we will demonstrate next, collisions due 
to bodies in resonance do not have a significant effect on the final shape of the radial dust distribution that we predict.

\begin{figure}
\begin{center}
    \includegraphics[width=0.5\textwidth]{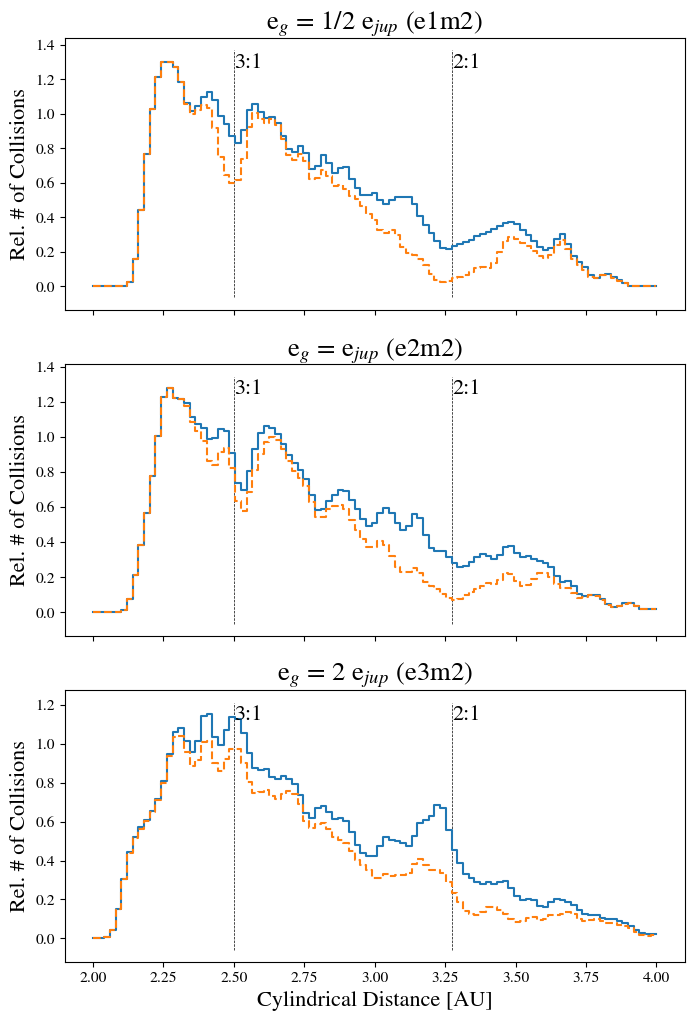}
    \caption{Here, collisions are instead ordered by cylindrical distance from the star. Features near the 3:1 and 2:1 MMRs are still present, but appear qualitatively different than in 
    semimajor axis space. At low eccentricities, a dip appears around the center of the resonance. At the highest eccentricity, a bump is formed instead. The dashed lines shown the 
    collision profile with collisions between bodies in resonance removed. The bump and dip features remain qualitatively the same, which suggests that they are produced by bodies 
    outside of the libration width of the resonance.\label{fig:coll_hist_r}}
\end{center}
\end{figure}

\section{Where Does the Dust End Up?}\label{sec:dust}

To construct a radial dust profile from the collision statistics, we begin by making two assumptions: (1) any dust generated by collisions is strongly coupled to the gas and (2) any 
subsequent spatial evolution of the collisional debris is insignificant. So long as both of these assumptions are true, we can use the radial locations of planetesimal collisions in the 
plane of the disk to generate a dust profile for each simulation.

We first provide a justification for assumption (1), which is as follows: At 3 au in the protosolar nebula, the mean free path of a gas particle is $\sim$ 50 cm (assuming a composition 
of pure hydrogen with a local volume density of $\rho_{g} = 3 \times 10^{-11}$ g cm$^{-3}$). This places any dust grains in the Epstein drag regime, with a stopping timescale given by

\begin{equation}\label{eq:ts_epstein}
    t_{s} = \frac{\rho s}{\rho_{g} v_{th}},
\end{equation}

\noindent where $\rho$ is the bulk density of the dust grain, $s$ is its size, and $v_{th}$ is the local thermal velocity of the gas. At 3 au in the protosolar nebula, $v_{th} \sim 10^{5}$ cm 
s$^{-1}$. Assuming a 1 mm dust grain with a $\rho$ = 2 g cm$^{-3}$, the stopping time is $\sim 10^{-4}$ yr.  This is orders of magnitude smaller than the 
orbital timescale; therefore, we conclude that collisionally generated dust grains will immediately couple to the gas.

Assumption (2) is less straightforward to justify and may not be reasonable in all cases. Although the spatial redistribution of the full-sized planetesimals and mm-sized dust grains 
is insignificant, the radial drift forces experienced by intermediate-sized bodies is much greater. Collisions between planetesimals do not immediately convert the mass to dust grains. 
Rather, the debris will undergo a collisional cascade process, which
gradually grinds the resulting bodies down to smaller sizes. A proper
treatment of the collisional cascade process is 
quite complicated and is well beyond the scope of this work. However, one may be able to ignore this process if the peak radial drift timescale is longer than the timescale for collisions 
between these bodies.

This occurs for bodies of the size at which $t_{s} \Omega$ = 1. In the Stokes drag regime at 3 au, this occurs for bodies of 1 meter in size. Following a similar 
calculation in \ref{sec:pl_drag} with a 1 m body, we find that an object of this size will drift across the 2:1 MMR on a timescale of $\sim$ 300 years. Although we do not model collisions 
between bodies of this size, a lower limit on the collision timescale for these objects can be obtained by extrapolating the planetesimal collision rate measured in our simulations down. For 
the e2m2 simulation, we find that 125 collisions occur within the libration width of the 2:1 MMR from T=2,000 yr to  T=5,000 yr. There are a total of 15,000 planetesimals within this region, 
which means that a single planetesimal undergoes a collision every $\sim$ 370,000 years on average. If the entire planetesimal population in this region were converted to 1 meter bodies 
with the same bulk density, the number density of bodies increases by a factor of $10^{15}$, while the geometric collision cross section decreases by a factor of $10^{10}$. The collision 
timescale would therefore decrease to $\sim$ 1 year, which is 2 orders of magnitude smaller than the drift timescale. As shown above, radial drift in the Stokes drag regime is maximized 
for bodies of 1 meter in size, which is only a factor of two larger than the mean free path of the gas particles. It is therefore not entirely clear if the Stokes drag law is appropriate here. 
Because of this, we also repeat the above calculation for bodies that experience Epstein drag. With this assumption, we find that radial drift is maximized for a cm-sized body and both the 
radial drift and collision timescale for bodies of that size decrease by a factor of 10 relative to the Stokes values. In both cases, collisions between maximally drifting bodies occur on a 
timescale two orders of magnitude shorter than the drift timescale. We therefore argue that it is reasonable and, indeed, best practice to start with such simple assumptions understand the 
physics before additional complexity is later added. We caution, however, that the collision timescale estimates provided here should be interpreted as lower limits.

Operating under these assumptions \footnote{As discussed in \cite{2017ApJ...850..103B}, the local generation of second-generation dust is only one 
consideration. The dust will also evolve spatially due to drag effects, but can also evolve in size or be re-accreted onto planetesimals. For simplicity, 
we focus on local dust sourcing only while acknowledging that subsequent evolution could \textbf{also} affect the morphology.}, a map of the relative concentration 
of second-generation dust can be constructed from the cylindrical distance at which the collisions occur. Here, cylindrical distance is defined as the separation between the central 
star and the location of collision in the ($r, \theta$) plane. This is shown in figure \ref{fig:coll_hist_r}. 
Similar to figure \ref{fig:coll_hist_a}, we use a KDE with a width of 0.02 au to assemble the collision locations into a radial distribution. Most 
strikingly, the bump that was present near the 2:1 resonance is no longer visible in cylindrical distance space, although it suddenly appears again in 
the e3m2 simulation. The local minimum that is visible near the 3:1 MMR also becomes a local maximum in the highest eccentricity case. To determine how the resonant 
bodies actually contribute to the radial collision profile, we excluded collisions that fall between $2.495 < a < 2.505$ au (near the 3:1 resonance) and 
between $3.2 < a < 3.35$ au (near the 2:1 resonance), which is shown by the dashed curve. As in figure \ref{fig:coll_hist_a}, the solid curves are normalized such that the area 
underneath is equal to 1. For the dashed curves, the normalization factor is scaled according to the number of collisions in the entire sample, compared to the number of collisions in the 
subsample, which excludes the collisions in resonance. Qualitatively, none of the bump or dip features present 
are removed by making this exclusion. This suggests that the features near resonance seen in semimajor axis space become smeared out in 
cylindrical distance space due to the large spread in eccentricity and orbital orientation.

\begin{figure*}
    \includegraphics[width=\textwidth]{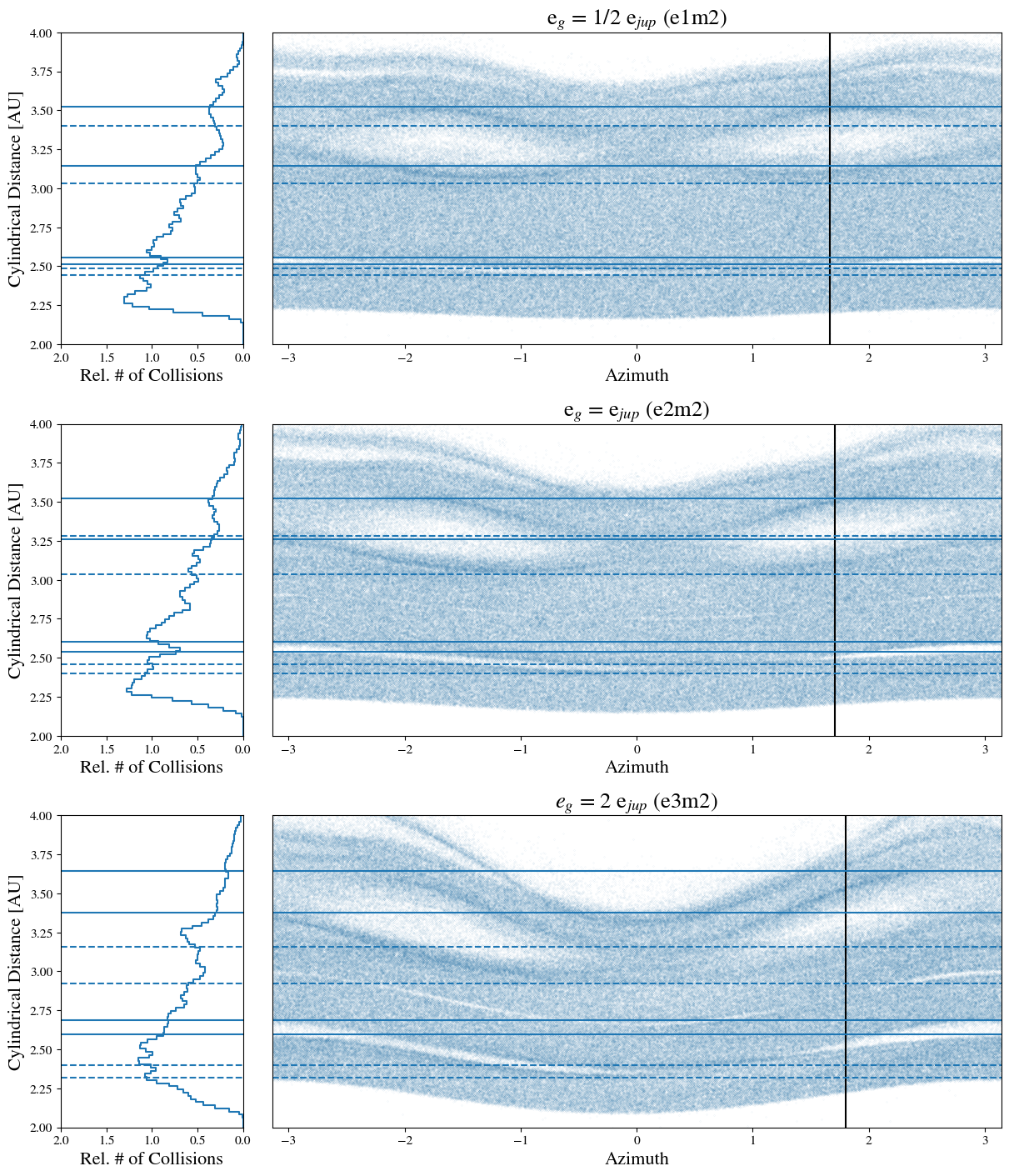}
    \caption{The collision profiles in figure \ref{fig:coll_hist_r} are shown alongside the positions of the planetesimals, in polar coordinates. The dashed and solid lines show the pericenter 
    and apocenter, respectively,  of bodies at each edge of the 3:1 and 2:1 MMR. In these figures, the longitude of pericenter of the planet lies at $\theta = 0$ and the present position of the 
    planet is indicated by the vertical line. For the 2:1 resonance, a bump, rather than a dip feature appears when the inner edge or the resonance's apocenter and outer edge's pericenter 
    distance cross. The collision profile near the 3:1 resonance does not appear to follow this same qualitative behavior.
    \label{fig:coll_polar_e}}
\end{figure*}

This also suggests that the prominent bumps and dips seen in figure \ref{fig:coll_hist_r} are mainly produced by bodies outside of the libration width of the resonance, although 
the resonant population does make some of the features more pronounced. To further understand this, we consider the nonaxisymmetric structure 
seen in figure \ref{fig:xy}. In figure \ref{fig:coll_polar_e}, we compare the radial collision profiles to the radial structure seen in the e1m2, e2m2 and 
e3m2 simulations. In the lowest forced eccentricity case (e1m2), the pileups near the edge of the 2:1 MMR line up with the boundaries of the dip 
feature seen in the radial collision profile. The locations of the edges of the resonances are calculated using equations \ref{eq:res_fo} and 
\ref{eq:res_so}. The pileups not as pronounced near the other resonances. As discussed previously, the planetesimal density enhancements and gaps 
seen in polar coordinates follow the position of the giant planet in its orbit, which is shown by the vertical line. For the e2m2 case, the density 
enhancements take up a much larger radial width, which smears out the corresponding collisional dust profile. In the highest forced eccentricity case 
(e3m2), the density enhancements occur well within an annulus containing the resonant region, completely altering the radial peak-valley dust 
morphology.

Although the nonaxisymmetric pileup of bodies near the edges of resonances has been seen previously 
\citep{2000Icar..143...45R, 2016ApJ...818..159T}, we could not find a satisfactory explanation for why it happens, which we will try to provide here. As 
mentioned in section \ref{sec:dynamics}, the circulation frequency of the critical angle slows as one approaches the edge of the resonance from the 
outside. Following the pendulum analogy, this is equivalent to
approaching the point where the pendulum becomes suspended in the
at top dead center. 
When the critical angle stays relatively stationary during an encounter, the net torque will drive the longitude of pericenter towards the point of 
conjunction \citep{1976ARA&A..14..215P}.  This is the basic mechanism that causes isolated resonances to be stable. Inside of resonance, the critical 
angle librates about an equilibrium value and this configuration is maintained. Just outside of resonance, however, the critical angle slowly drifts away, with the 
drift direction changing sign across the location of exact resonance. The net result of this is that bodies just outside of the resonance experience a torque from 
the planet that temporarily drives their pericenters toward the current orbital position of the planet. After the encounter ends, differential rotation of the disc 
slowly undoes the configuration. As the planet again passes by, adjacent planetesimals near the resonance edges temporarily align 
themselves in this configuration. This explains why the concentration of planetesimals appears to `follow' the planet. Connecting this back to the radial collision profiles shown in 
figure \ref{fig:coll_hist_r}, these bodies at the edges of the resonances must be responsible for the bumps and dips morphology seen because excluding collisions between bodies inside 
of the MMRs does not qualitatively alter the profiles.

\begin{figure*}
    \includegraphics[width=\textwidth]{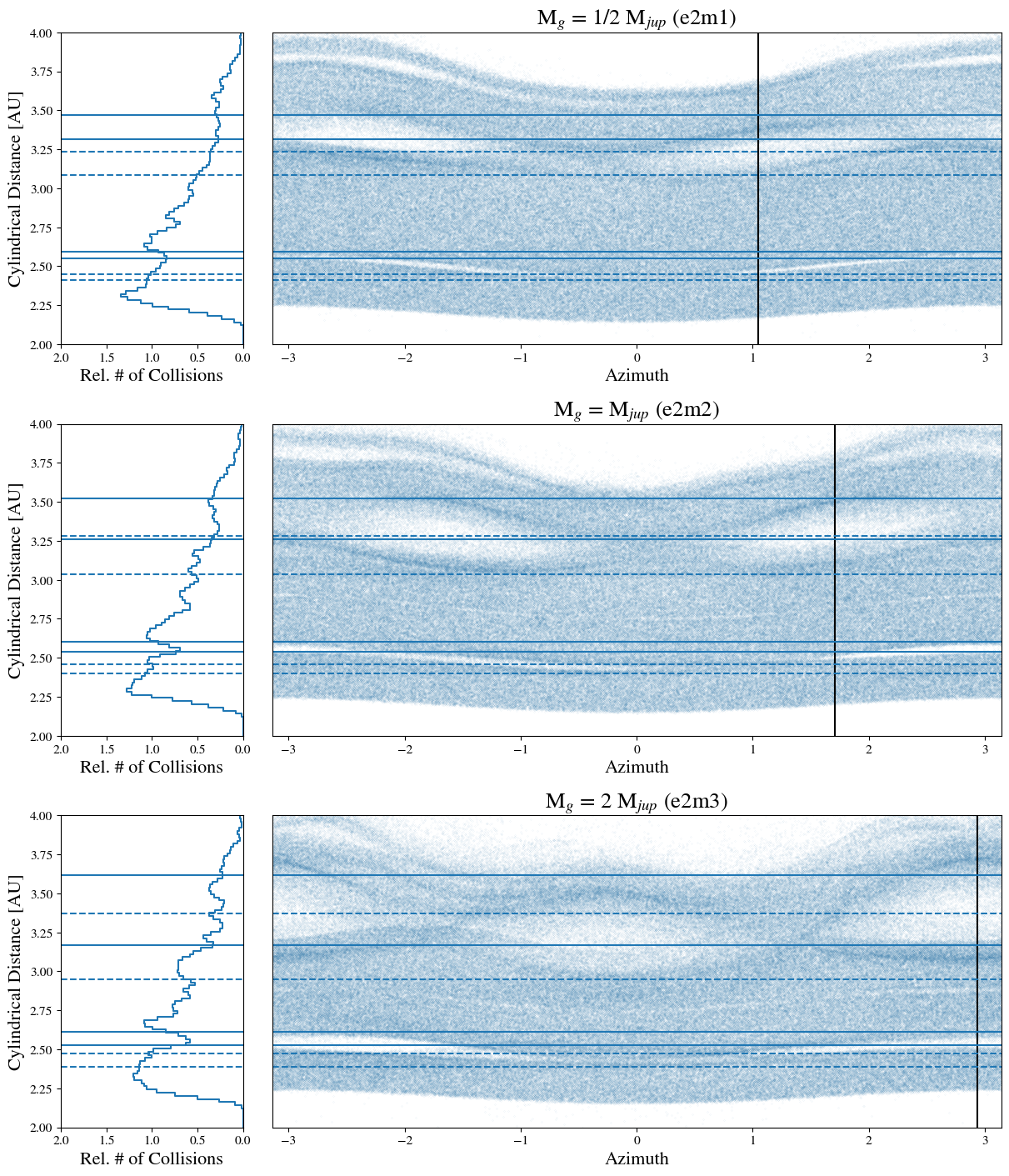}
    \caption{Similar to figure \ref{fig:coll_polar_e}, except the eccentricity of the planet is kept constant and the mass is varied.
    This has the effect of changing the width of the resonances, without altering the relative apo or peri distances of
    bodies near the edges. Except for the highest mass case, the apocenter and pericenter distances of the inner and outer edges of the 2:1 
    resonance are too close together to produce much of a dip or a bump feature.\label{fig:coll_polar_m}}
\end{figure*}

\citet{2016ApJ...818..159T} provide a different explanation for this phenomenon, suggesting that the nonaxisymmetric structure is the product 
of the path that a resonant test particle takes in a frame co-rotating with the planet (see figure 8.4 of \citet{1999ssd..book.....M}). In other words, the gap structure is argued to be 
due to the interfaces between the low eccentricity, non-resonant and high eccentricity, resonant regions of the disc. This explanation, 
however, does not hold for a large collection of planetesimals, especially when there is no forced eccentricity. In such a case, the orbits of 
planetesimals near resonance would be randomly aligned, and the axisymmetric structure should become washed out. Furthermore, when the perturbing 
planet is on a circular orbit, this structure is still present (see figure 3 of \citet{2016ApJ...818..159T} and Appendix \ref{sec:boley_plot}). The only 
connection between the structure of the gaps in the disc and the path of a resonant particle in the corotating frame is that both phenomena produce the 
same symmetries, which depend on the particular MMR considered.

With no secular forcing, this dynamical phenomena should produce an underdensity in the radial surface density profile near the center of the 
resonance and an overdensity at each edge in cylindrical distance space. When a forced eccentricity is introduced, the radial location of the 
resonance edges vary over the course of an orbit. The maximum and minimum radial distance that the resonance edges occupy is indicated by the 
solid and dashed lines in figure \ref{fig:coll_polar_e}. If the the aphelion distance of the inner edge becomes close to the perihelion distance of 
the outer edge, the overdense regions on each side of the resonance meet and we should expect to see the central dip feature in the collision profile 
disappear. This is exactly what appears to be happening in the middle panel of figure \ref{fig:coll_polar_e}. As the forced eccentricity is increased 
further, a region forms at the center of the resonance where planetesimals on both sides of the resonance spend time (although not simultaneously). 
When this occurs, a bump feature forms in the collision profile near the center of the resonance. This is apparent in the bottom panel of figure 
\ref{fig:coll_polar_e}.

Although the same phenomenon appears to be happening near the 3:1 MMR, the pileups at the edges of the resonance are not as well defined. We 
attribute this to the more complex symmetry, compared to the 2:1 MMR. In addition, a number of other nearby resonances, including the 7:2 (at 
2.25 au), the 10:3 (at 2.33 au), the 8:3 (at 2.71 au) and the 5:2 (at 2.82 au) also likely are contributing to the collision profile in cylindrical distance 
space near the 3:1 MMR. The 3:1 MMR is also much narrower than the 2:1, which means that any dip or bump features produced by it would 
require much higher resolution observations. For these reasons, we will focus on the 2:1 MMR as the main diagnostic indicator.

\subsection{Varying the Mass}

The dynamical effects of varying the mass of the planet are somewhat simpler, in that doing so does not affect the forced eccentricity of the 
planetesimal disc. Instead, only the width of the resonances changes. The width of a first order resonance scales with $m$ (see equation 
\ref{eq:res_so}), while the leading order terms in the resonant part of the disturbing function, which set the strength of the resonance, also scale as 
$m$. For the 2:1 MMR, the dynamics near the resonance are equally sensitive to changes in eccentricity and mass.

We show the polar structure of the e2m1, e2m2 and e2m3 simulations in figure \ref{fig:coll_polar_m} alongside the radial collision profile. In all three 
cases, the eccentricity of the perturbing planet is set to $e_{jup}$. Changes in the apocenter and pericenter distances of the edges of the 
resonances are entirely due to changes in the resonance width. For the e2m1 and e2m2 simulations, the inner apocenter and outer pericenter 
distances near the 2:1 MMR are quite similar and no strong features appear in the collision profile near this region. For the e2m3 case, the edges of
the resonances are sufficiently separated to allow what appears to be the beginning of gap to form near the center of the 2:1 resonance in the collision profile.

\subsection{Observability of the Dust}

While the collisional dust profiles show radial amplitude variations, we still need to assess whether those variations could be observable. To proceed, 
we create a sky model for simulated ALMA observations as follows. First, we use the radially-averaged collision profile from each simulation (see figures 
\ref{fig:coll_polar_e} and \ref{fig:coll_polar_m}) as the template for an azimuthally symmetric disc. We then scale the size of each profile by a factor of 
ten, which places the perturbing planet at 52 au. This scaling is permitted by the dynamics (both the resonance widths and the forced eccentricities 
are scale-free) and makes the disc comparable in size to the many discs that have now been observed by ALMA \citep{2018ApJ...869L..42H}. The 
angular size scale is then set by envisaging a face-on disc at a distance of 100 pc.  

The sky model intensity (flux per pixel) is produced by interpolating the radial collision frequency onto a 2D Cartesian grid with cell widths of 
approximately 2 mas. In doing so, we choose to make the intensity proportional to the collision frequency (i.e., the dust) because it is the most 
straightforward case. Many different parameterizations are possible, depending on optical depth, grain size, disc temperature, etc., but we will avoid 
such complications here, as they would affect the overall brightness profile, but not the presence of any gaps or rings. We normalize the intensity by
setting the total disc flux density to be 100 mJy at about 350 GHz.  

The simulated observations are performed using the {\sc simobserve} task in {\sc CASA} \citep{2007ASPC..376..127M}.  The disc is given an 
$\rm RA = 11^{\rm h} 01^{\rm m} 52^{\rm s}$ and a $\delta = -34^\circ 42\arcmin 17\arcsec$ (i.e., the J2000 coordinates for TW Hydrae) and is 
``observed'' through transit on March 20, 2020. We observe the disc using configurations 8, 9, and 10 at 350 GHz (cycle 6 antenna file), spending six 
hours on-source in each configuration\footnote{We have imagined that the TAC really likes us.}.  The visibilities are corrupted with thermal noise, 
setting the precipitable water vapor to 1.5 mm. They are then combined, imaged, and cleaned
using the task {\tt tclean}. Imaging uses Briggs weighting with a robustness parameter of -1. 

\begin{figure}
\begin{center}
    \includegraphics[width=0.5\textwidth]{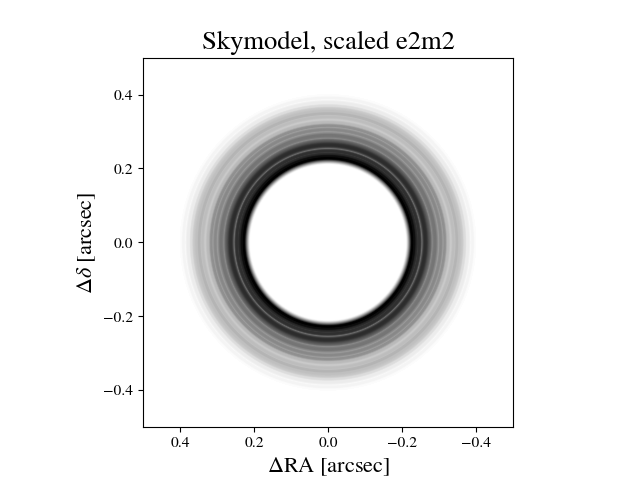}
    \includegraphics[width=0.5\textwidth]{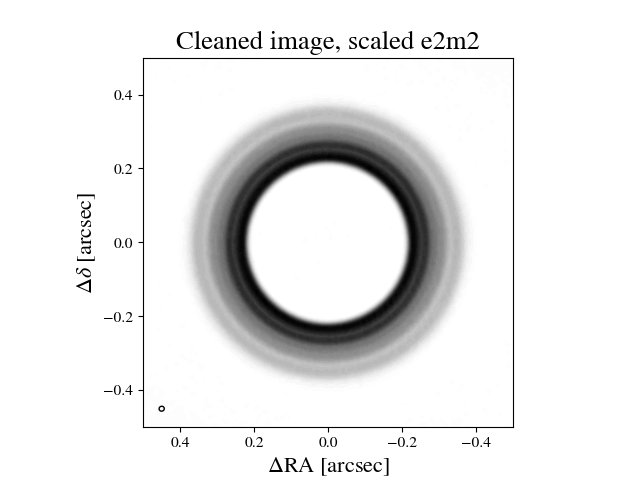}
    \caption{Simulated ALMA observations for the nominal (e2m2) case, operating under the assumption that the dust profile closely traces the planetesimal collision profile.  Top: 
    The sky model based on the radial collision distribution. The size of the disc has been scaled by a factor of 10 and placed at  a distance of 100 pc. Bottom: The cleaned image using 
    combined observations in configurations 8, 9, and 10. The circle in the bottom left indicates the simulated beam size. Gaps in the dust due to the 3:1 and 2:1 resonances are visible. 
    \label{fig:alma_sim_obs}}
\end{center}
\end{figure}

The results for the nominal simulation (e2m2) are shown in figure \ref{fig:alma_sim_obs}. The 3:1 and 2:1 resonances can be easily identified in the 
cleaned image and in the corresponding radial profile.  Moreover, the features are qualitatively similar to the bright rings and dark gaps seen in actual 
disc profiles \citep{2015ApJ...808L...3A}.

\begin{figure*}
    \includegraphics[width=\textwidth]{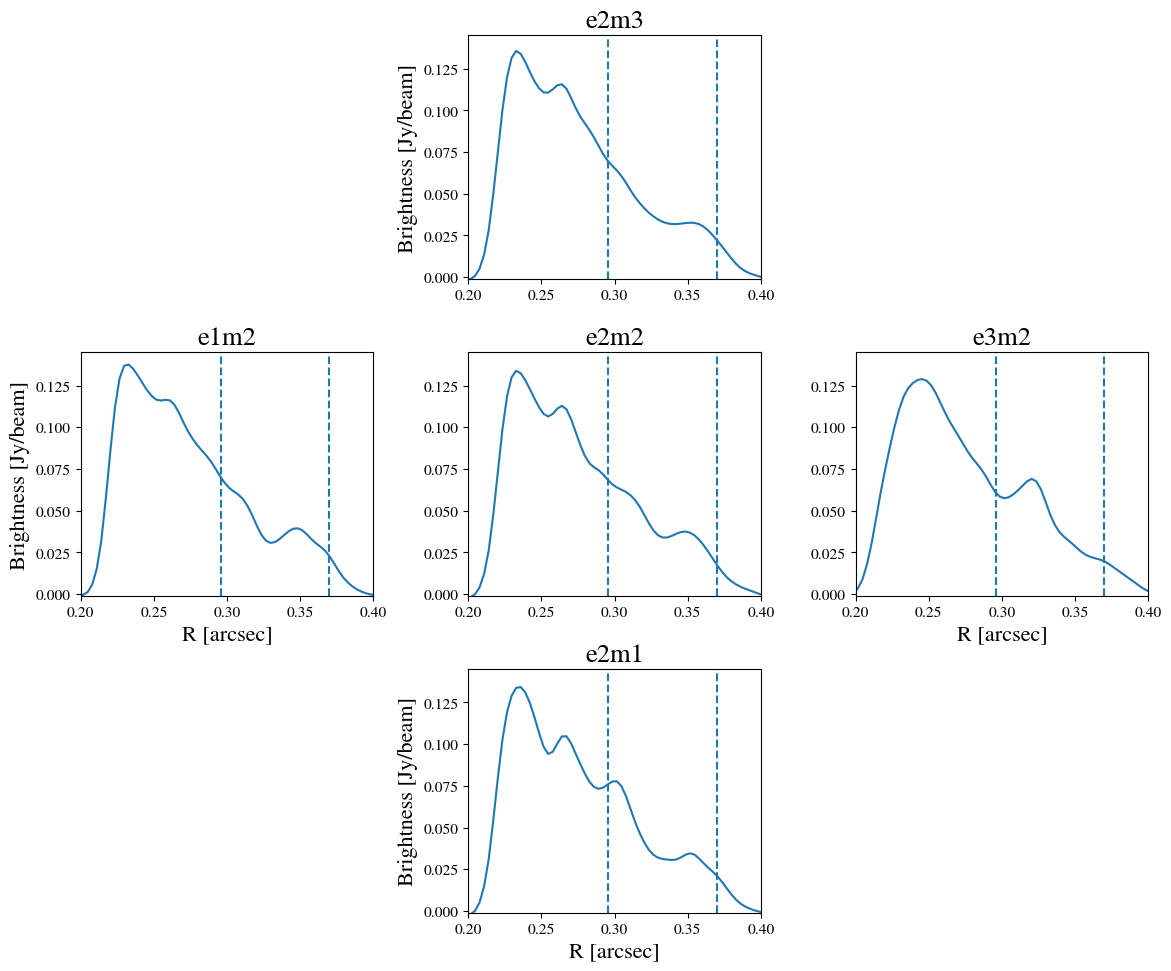}
    \caption{The azimuthally-averaged radial profile based on the simulated cleaned images from all five simulations. The eccentricity of the perturbing planet increases from left to right, 
    while the mass of the planet increases from bottom to top. The vertical dashed lines indicate the locations of the 3:1 and 2:1 resonances. As in figure \ref{fig:alma_sim_obs}, the size 
    scale of the disc has been expanded by a factor of 10 and then placed at a distance of 100 pc.\label{fig:alma_profiles}}
\end{figure*}

In figure \ref{fig:alma_profiles}, we show the 
azumithally-averaged radial profiles constructed from the simulated cleaned image from all five 
simulations. In all cases, a bump or a dip feature is clearly visible at the location of the 3:1 and 2:1 MMRs, indicated by the dashed vertical lines. 
As mentioned previously, the feature at the 2:1 MMR acts as the actual diagnostic indicator, while the 3:1 resonance (along with the gap that would presumably 
open at the location of the planet) is mainly useful for 
determining where the 2:1 MMR is actually located.

\section{Constraining the Mass and Eccentricity of the Planet}\label{sec:constrain}

\begin{figure}
\begin{center}
    \includegraphics[width=0.5\textwidth]{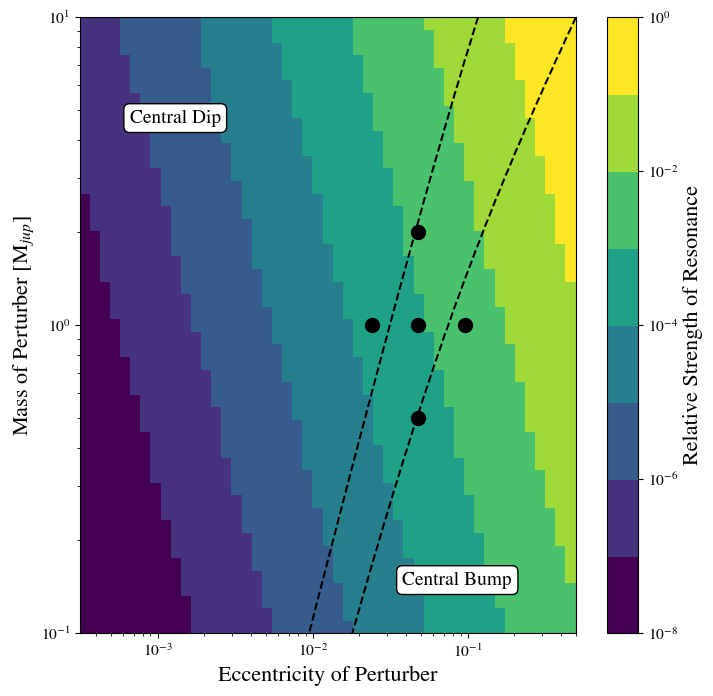}
    \caption{The presence of a dip or bump in the collision profile (and therefore the dust profile) near the 2:1 MMR can be used to constrain the mass and eccentricity
    of the perturbing planet. The color scale indicates the relative strength of the features, while the dashed lines
    indicate boundaries in parameter space where a dip or a bump will be produced at the resonance. Combinations of mass and eccentricity that fall 
    between the dashed lines correspond to an inner apocenter and outer pericenter separation at the edges of the resonance that is smaller than the 
    resonance width and therefore will not create any significant feature in the collision profile.\label{fig:bump_dip_diag}}
\end{center}
\end{figure}

The simple bump vs dip structure that we expect to reveal itself in the dust emission from colliding planetesimals in near-resonance with a giant 
planet could potentially be used to place constraints on the mass and eccentricity of the planet. Near the 2:1 MMR, the central dip feature is only 
produced when bodies at the edges of the resonance stay sufficiently separated in cylindrical distance over the course of an orbit. This is achieved 
when either the resonance width is large or the forced eccentricity is small. The bump feature, on the other hand, is produced when there is a 
sufficiently large amount of overlap between the apocenters and pericenters of the inner and outer edges of the resonance, respectively. This is 
achieved when the resonance is narrow and the forced eccentricity is large.

In figure \ref{fig:bump_dip_diag}, we show the constraints that the presence of a bump or a dip or dip at the 2:1 MMR places on possible values of $(m_{g}, e_{g})$. The dashed
lines indicate regions of parameter space where qualitatively different features in the collision profile are expected to form. Above the upper dashed line, the 
resonance is wide while the radial excursion distance of planetesimals at the edge of the resonance is small, which gives rise to a dip in the collision profile. 
Below the lower dashed line, the resonance is narrow, while the radial excursion distance is large, allowing the apocenter and pericenter distances of the edges 
of the resonance to overlap and produce a bump in the collision profile. Between the dashed lines, the absolute value of separation between the apocenter and 
pericenter distances of the inner and outer edges of the resonance is less than the resonance width. In this region of parameter space, no significant features in 
the collision profile near the 2:1 MMR are expected to form. 

Because the strength\footnote{More specifically, the strength of the resonant term in the disturbing potential} of the 2:1 resonance scales with $e_{g}$ and  $m_{g}$, a sufficiently low 
mass and low eccentricity planet will not create enough of a perturbation in the planetesimal disc to produce a detectable bump or dip feature. The colored contours in figure 
\ref{fig:bump_dip_diag} indicate the relative strength of the resonance as these quantities are varied. Although the resonance strength is equally sensitive to changes in mass and 
eccentricity, the peri-apocenter overlap of the resonance edges is more sensitive to changes in eccentricity. This is because changes to eccentricity affect both the resonance width and 
the forced eccentricity. As we showed in figure \ref{fig:coll_polar_m}, changing $m_{g}$ by a factor of 4 produced very minimal changes to the resulting dust profile.

In order to actually identify the 2:1 MMR in the dust emission, at least one other resonance must be visible. Assuming the gravitational field in the 
disc is near-Keplerian, the distance ratios between two features can be used to determine period ratios, which can be used to confirm whether the 
features seen are indeed resonances. Although the 3:1 MMR does not appear to follow the simple bump vs dip dichotomy described above, its 
presence should be marginally detectable in all of the cases shown above.

We would again like to emphasize that the dust emission profiles in figure \ref{fig:alma_profiles} are constructed upon the assumption that the radial dust structure closely traces the 
collision profile of the planetesimals. For this assumption to be valid, the resulting dust must remain locally confined, which can be achieved if the grains are reaccreted onto 
planetesimals \citep{2015SciA....1E0109J}. It is also necessary that the collisional cascade from which the dust grains form plays before radial drift has a chance to move debris away 
from the resonances. Although we showed with a back-of-the-envelope calculation (see section \ref{sec:dust}) that this is plausible for a typical protoplanetary disk with properties similar 
to the protosolar nebula, a more thorough treatment of collisional grinding and the subsequent evolution of the debris is necessary to be able to interpret the dust emission profiles 
presented here as anything more than upper limits.

\section{Summary and Conclusions}\label{sec:conclusions}

In this work, we have shown that mean-motion resonances with a perturbing planet produce significant local variations in the collision rate of a 
planetesimal disc. In contrast to \citet{2000Icar..143...45R}, we find that the more prominent interior MMRs, including the 2:1, 3:1 and 5:3, all 
produce structure in the collision profile as a function of semimajor axis. Furthermore, we find that a series of distinctly different features appear 
when collisions are ordered by cylindrical distance from the central star. The morphology of these features are tied to dynamical behavior of bodies near the edges of the 
resonances, where the circulation frequency of the critical angle approaches infinity. Particularly near the 2:1 MMR, we find that a bump or dip in the collision profile forms when collisions 
are ordered by cylindrical distance. The presence of one of these two features depends on the peri- and apocenter distances of the edges of the resonance, relative to the libration width. 
Because these quantities depend on the mass and eccentricity of the perturbing planet, these properties of the planet are actually encoded in the collision profile.

Near the interior 2:1 MMR, we find that a distinct bump or dip feature is generated in the collision profile, depending on the properties of the perturbing 
planet. The presence of one of these two structures can used to constrain the mass and eccentricity of the planet. For a high mass, low eccentricity 
planet, a dip will form because the edges of the resonance, where many collisions occur, stay sufficiently separated. If the planet 
has a low enough mass (which shrinks the size of the resonance) or is sufficiently eccentric (which decreases the separation between the apocenter 
and pericenter distances of the edges of the resonance), a bump will instead form. We tested this hypothesis for five different combinations of the 
planet's mass and eccentricity and found that a distinct bump or dip signature is produced as long as the planet properties are sufficiently far from 
the dividing line in parameter space. This diagnostic is more useful for massive, eccentric planets, because the strength of the resonant 
perturbations scale linearly with both of these quantities.

Although the planetesimal collision profile would not be directly observable in a planet-forming disc, the dust from the resulting collisional cascade could potentially be used to trace 
it. So long as the dust grains are well-coupled to the gas and radial drift does not have a significant effect on the morphology of the dust, the bump or dip features seen in the collision 
profile will also appear in the dust profile. Assuming that this is the case, we have generated simulated observations of a protoplanetary disk with ALMA and showed that the bump or dip 
features seen in the collision profiles should be detectable through the dust emission. From a dust emission profile with sufficiently strong radial features, one could constrain the 
properties of a perturbing planet on an exterior orbit in the following way: (1) identify the locations of two MMRs in the disc (2) measure the spacing between them to verify that one of the 
features is indeed associated with the 2:1 MMR (3) determine whether an under- or overdensity in the radial dust profile exists near this location (4) use the presence of one of these two 
features to determine the region of the eccentricity-mass plane in which the planet lies.

In line with \citet{2017ApJ...850..103B}, the results of this work suggest that, in planet forming discs, the observed dust structure should not be 
interpreted as simply the result of primordial dust that is perturbed by the planets and gas. Instead, some morphological features could be the product 
of collisional dynamics between larger bodies. Within the confines of the simulations presented here, we have shown that this second-generation dust 
can give rise to significant features in the dust emission profile. For this reason, we emphasize that planet-forming discs should be thought of as 
existing along an evolutionary continuum, with the youngest discs being dominated by primordial dust and gas and the most evolved discs being 
comprised of mainly collisionally generated dust.

The results presented here appear to be broadly consistent with the planetesimal dynamics seen by \citet{2016ApJ...818..159T}. In their case, a dip 
feature can be seen in the azimuthally averaged surface density profile, although they did not test a high enough eccentricity planet in any of their 
simulations to produce a bump. Another thing to note is that the gap morphology seen by \citet{2016ApJ...818..159T} was markedly different for the 
2:1 exterior MMR. Whether this would alter the radial collision profile for the exterior, rather than the interior 2:1 resonance, is not immediately clear.  
On one hand, the solid surface density diminishes with distance ($\sim r^{-3/2}$ for the MMSN \citep{1981PThPS..70...35H}), which would weaken the signal from locally 
generated collisional dust. On the other hand, exterior MMRs are quite effective at capturing inward drifting bodies \citep{1985Icar...62...16W}, which could locally enhance the 
surface density well beyond the MMSN value. A full treatment of the collisional cascade process and the radial drift of debris is a subject that we leave for future work.

Compared to using the corotating gap opened by the planet as a diagnostic \citep{2013ApJ...777L..31D, 2016ApJ...820...29D}, measuring 
variations in the dust emission near MMRs is much more subtle and requires much higher spatial resolution and sensitivity. As we have shown, the 
bump vs dip feature near the 2:1 MMR is marginally detectable with ALMA for the nearest protoplanetary discs, so long as the dust profile closely matches the collisional profile of 
the planetesimals. Another complicating factor is that 
the inner $\sim$ 10 au of most protoplanetary discs are optically thick in the sub-mm. The solution to this problem is to instead observe at longer 
wavelengths where the inner disc is optically thin. This comes at the expense of much poorer resolution. However, future radio facilities like the NG-
VLA are expected to achieve sub-au resolution for nearby planet-forming discs \citep{2018ASPC..517..147R}. In the more near-term, techniques 
like Gaussian process fitting present a promising way to recover substructure at sub-beam resolution \citep{2020arXiv200507709J}.

\section*{Acknowledgements}
This work was facilitated through the use of advanced computational, storage and networking infrastructure provided by the Hyak supercomputer system at the 
University of Washington. We acknowledge the people of the Dkhw’Duw’Absh, the Duwamish Tribe, the Muckleshoot Tribe, and other tribes on whose traditional 
lands we have performed this work.

\textit{Software:} Astropy \citep{2013A&A...558A..33A}, {\sc CASA} \citep{2007ASPC..376..127M}, {\sc ChaNGa} \citep{2008IEEEpds...ChaNGa, 2015AphCom..2..1}, Matplotlib 
\citep{2007CSE.....9...90H}, NumPy \citep{2011CSE....13b..22V}, Pandas \citep{mckinney-proc-scipy-2010}, {\sc PYNBODY} \citep{2013ascl.soft05002P}, {\sc sklearn} \citep{scikit-learn}

\section*{Data Availability}
The data presented in this article are available on our GitHub repository, at \url{https://github.com/spencerw/planetesimal_coll_paper}.

\bibliographystyle{mnras}
\bibliography{main}

\begin{thebibliography}{}
\makeatletter
\relax
\def\mn@urlcharsother{\let\do\@makeother \do\$\do\&\do\#\do\^\do\_\do\%\do\~}
\def\mn@doi{\begingroup\mn@urlcharsother \@ifnextchar [ {\mn@doi@}
  {\mn@doi@[]}}
\def\mn@doi@[#1]#2{\def\@tempa{#1}\ifx\@tempa\@empty \href
  {http://dx.doi.org/#2} {doi:#2}\else \href {http://dx.doi.org/#2} {#1}\fi
  \endgroup}
\def\mn@eprint#1#2{\mn@eprint@#1:#2::\@nil}
\def\mn@eprint@arXiv#1{\href {http://arxiv.org/abs/#1} {{\tt arXiv:#1}}}
\def\mn@eprint@dblp#1{\href {http://dblp.uni-trier.de/rec/bibtex/#1.xml}
  {dblp:#1}}
\def\mn@eprint@#1:#2:#3:#4\@nil{\def\@tempa {#1}\def\@tempb {#2}\def\@tempc
  {#3}\ifx \@tempc \@empty \let \@tempc \@tempb \let \@tempb \@tempa \fi \ifx
  \@tempb \@empty \def\@tempb {arXiv}\fi \@ifundefined
  {mn@eprint@\@tempb}{\@tempb:\@tempc}{\expandafter \expandafter \csname
  mn@eprint@\@tempb\endcsname \expandafter{\@tempc}}}

\bibitem[\protect\citeauthoryear{{ALMA Partnership} et~al.,}{{ALMA Partnership}
  et~al.}{2015}]{2015ApJ...808L...3A}
{ALMA Partnership} et~al., 2015, \mn@doi [\apj] {10.1088/2041-8205/808/1/L3},
  \href {https://ui.adsabs.harvard.edu/abs/2015ApJ...808L...3A} {808, L3}

\bibitem[\protect\citeauthoryear{{Adachi}, {Hayashi}  \& {Nakazawa}}{{Adachi}
  et~al.}{1976}]{1976PThPh..56.1756A}
{Adachi} I.,  {Hayashi} C.,   {Nakazawa} K.,  1976, \mn@doi [Progress of
  Theoretical Physics] {10.1143/PTP.56.1756}, \href
  {https://ui.adsabs.harvard.edu/abs/1976PThPh..56.1756A} {56, 1756}

\bibitem[\protect\citeauthoryear{{Andrews} et~al.,}{{Andrews}
  et~al.}{2016}]{2016ApJ...820L..40A}
{Andrews} S.~M.,  et~al., 2016, \mn@doi [\apj] {10.3847/2041-8205/820/2/L40},
  \href {https://ui.adsabs.harvard.edu/abs/2016ApJ...820L..40A} {820, L40}

\bibitem[\protect\citeauthoryear{{Astropy Collaboration} et~al.,}{{Astropy
  Collaboration} et~al.}{2013}]{2013A&A...558A..33A}
{Astropy Collaboration} et~al., 2013, \mn@doi [\aap]
  {10.1051/0004-6361/201322068}, \href
  {https://ui.adsabs.harvard.edu/abs/2013A&A...558A..33A} {558, A33}

\bibitem[\protect\citeauthoryear{{Augereau}, {Nelson}, {Lagrange}, {Papaloizou}
   \& {Mouillet}}{{Augereau} et~al.}{2001}]{2001A&A...370..447A}
{Augereau} J.~C.,  {Nelson} R.~P.,  {Lagrange} A.~M.,  {Papaloizou} J.~C.~B.,
  {Mouillet} D.,  2001, \mn@doi [\aap] {10.1051/0004-6361:20010199}, \href
  {https://ui.adsabs.harvard.edu/abs/2001A&A...370..447A} {370, 447}

\bibitem[\protect\citeauthoryear{{Barnes} \& {Hut}}{{Barnes} \&
  {Hut}}{1986}]{1986Natur.324..446B}
{Barnes} J.,  {Hut} P.,  1986, \mn@doi [\nat] {10.1038/324446a0}, \href
  {https://ui.adsabs.harvard.edu/abs/1986Natur.324..446B} {324, 446}

\bibitem[\protect\citeauthoryear{{Boley}}{{Boley}}{2017}]{2017ApJ...850..103B}
{Boley} A.~C.,  2017, \mn@doi [\apj] {10.3847/1538-4357/aa93d7}, \href
  {https://ui.adsabs.harvard.edu/\#abs/2017ApJ...850..103B} {850, 103}

\bibitem[\protect\citeauthoryear{{Brouwer} \& {Clemence}}{{Brouwer} \&
  {Clemence}}{1961}]{1961mcm..book.....B}
{Brouwer} D.,  {Clemence} G.~M.,  1961, {Methods of celestial mechanics}.
Academic Press

\bibitem[\protect\citeauthoryear{{Cieza} et~al.,}{{Cieza}
  et~al.}{2016}]{2016Natur.535..258C}
{Cieza} L.~A.,  et~al., 2016, \mn@doi [\nat] {10.1038/nature18612}, \href
  {https://ui.adsabs.harvard.edu/abs/2016Natur.535..258C} {535, 258}

\bibitem[\protect\citeauthoryear{{Dipierro}, {Price}, {Laibe}, {Hirsh},
  {Cerioli}  \& {Lodato}}{{Dipierro} et~al.}{2015}]{2015MNRAS.453L..73D}
{Dipierro} G.,  {Price} D.,  {Laibe} G.,  {Hirsh} K.,  {Cerioli} A.,   {Lodato}
  G.,  2015, \mn@doi [\mnras] {10.1093/mnrasl/slv105}, \href
  {https://ui.adsabs.harvard.edu/abs/2015MNRAS.453L..73D} {453, L73}

\bibitem[\protect\citeauthoryear{{Dobinson}, {Leinhardt}, {Dodson-Robinson}  \&
  {Teanby}}{{Dobinson} et~al.}{2013}]{2013ApJ...777L..31D}
{Dobinson} J.,  {Leinhardt} Z.~M.,  {Dodson-Robinson} S.~E.,   {Teanby} N.~A.,
  2013, \mn@doi [\apj] {10.1088/2041-8205/777/2/L31}, \href
  {https://ui.adsabs.harvard.edu/abs/2013ApJ...777L..31D} {777, L31}

\bibitem[\protect\citeauthoryear{{Dobinson}, {Leinhardt}, {Lines}, {Carter},
  {Dodson-Robinson}  \& {Teanby}}{{Dobinson}
  et~al.}{2016}]{2016ApJ...820...29D}
{Dobinson} J.,  {Leinhardt} Z.~M.,  {Lines} S.,  {Carter} P.~J.,
  {Dodson-Robinson} S.~E.,   {Teanby} N.~A.,  2016, \mn@doi [\apj]
  {10.3847/0004-637X/820/1/29}, \href
  {https://ui.adsabs.harvard.edu/abs/2016ApJ...820...29D} {820, 29}

\bibitem[\protect\citeauthoryear{{Dong}, {Li}, {Chiang}  \& {Li}}{{Dong}
  et~al.}{2018}]{2018ApJ...866..110D}
{Dong} R.,  {Li} S.,  {Chiang} E.,   {Li} H.,  2018, \mn@doi [\apj]
  {10.3847/1538-4357/aadadd}, \href
  {https://ui.adsabs.harvard.edu/abs/2018ApJ...866..110D} {866, 110}

\bibitem[\protect\citeauthoryear{{Gaspar} \& {Rieke}}{{Gaspar} \&
  {Rieke}}{2020}]{2020PNAS..117.9712G}
{Gaspar} A.,  {Rieke} G.,  2020, \mn@doi [Proceedings of the National Academy
  of Science] {10.1073/pnas.1912506117}, \href
  {https://ui.adsabs.harvard.edu/abs/2020PNAS..117.9712G} {117, 9712}

\bibitem[\protect\citeauthoryear{{Hayashi}}{{Hayashi}}{1981}]{1981PThPS..70...35H}
{Hayashi} C.,  1981, \mn@doi [Progress of Theoretical Physics Supplement]
  {10.1143/PTPS.70.35}, \href
  {http://adsabs.harvard.edu/abs/1981PThPS..70...35H} {70, 35}

\bibitem[\protect\citeauthoryear{{Huang} et~al.,}{{Huang}
  et~al.}{2018}]{2018ApJ...869L..42H}
{Huang} J.,  et~al., 2018, \mn@doi [\apjl] {10.3847/2041-8213/aaf740}, \href
  {https://ui.adsabs.harvard.edu/abs/2018ApJ...869L..42H} {869, L42}

\bibitem[\protect\citeauthoryear{{Hunter}}{{Hunter}}{2007}]{2007CSE.....9...90H}
{Hunter} J.~D.,  2007, \mn@doi [Computing in Science and Engineering]
  {10.1109/MCSE.2007.55}, \href
  {https://ui.adsabs.harvard.edu/abs/2007CSE.....9...90H} {9, 90}

\bibitem[\protect\citeauthoryear{{Ida} \& {Makino}}{{Ida} \&
  {Makino}}{1992}]{1992Icar...96..107I}
{Ida} S.,  {Makino} J.,  1992, \mn@doi [\icarus]
  {10.1016/0019-1035(92)90008-U}, \href
  {https://ui.adsabs.harvard.edu/abs/1992Icar...96..107I} {96, 107}

\bibitem[\protect\citeauthoryear{{Ida}, {Kokubo}  \& {Makino}}{{Ida}
  et~al.}{1993}]{1993MNRAS.263..875I}
{Ida} S.,  {Kokubo} E.,   {Makino} J.,  1993, \mn@doi [\mnras]
  {10.1093/mnras/263.4.875}, \href
  {https://ui.adsabs.harvard.edu/abs/1993MNRAS.263..875I} {263, 875}

\bibitem[\protect\citeauthoryear{Isella et~al.,}{Isella
  et~al.}{2016}]{PhysRevLett.117.251101}
Isella A.,  et~al., 2016, \mn@doi [Phys. Rev. Lett.]
  {10.1103/PhysRevLett.117.251101}, 117, 251101

\bibitem[\protect\citeauthoryear{{Jennings}, {Booth}, {Tazzari}, {Rosotti}  \&
  {Clarke}}{{Jennings} et~al.}{2020}]{2020arXiv200507709J}
{Jennings} J.,  {Booth} R.~A.,  {Tazzari} M.,  {Rosotti} G.~P.,   {Clarke}
  C.~J.,  2020, arXiv e-prints, \href
  {https://ui.adsabs.harvard.edu/abs/2020arXiv200507709J} {p. arXiv:2005.07709}

\bibitem[\protect\citeauthoryear{{Jetley}, {Gioachin}, {Mendes}  \&
  {Quinn}}{{Jetley} et~al.}{2008}]{2008IEEEpds...ChaNGa}
{Jetley} P.,  {Gioachin} F.,  {Mendes} C.~{Kale} L.,   {Quinn} T.,  2008,
  Proceedings of IEEE International Parallel and Distributed Processing
  Symposium

\bibitem[\protect\citeauthoryear{{Johansen}, {Mac Low}, {Lacerda}  \&
  {Bizzarro}}{{Johansen} et~al.}{2015}]{2015SciA....1E0109J}
{Johansen} A.,  {Mac Low} M.-M.,  {Lacerda} P.,   {Bizzarro} M.,  2015, \mn@doi
  [Science Advances] {10.1126/sciadv.1500109}, \href
  {https://ui.adsabs.harvard.edu/abs/2015SciA....1E0109J} {1, 1500109}

\bibitem[\protect\citeauthoryear{{Kokubo} \& {Ida}}{{Kokubo} \&
  {Ida}}{2002}]{2002ApJ...581..666K}
{Kokubo} E.,  {Ida} S.,  2002, \mn@doi [\apj] {10.1086/344105}, \href
  {https://ui.adsabs.harvard.edu/abs/2002ApJ...581..666K} {581, 666}

\bibitem[\protect\citeauthoryear{{Leinhardt} \& {Richardson}}{{Leinhardt} \&
  {Richardson}}{2005}]{2005ApJ...625..427L}
{Leinhardt} Z.~M.,  {Richardson} D.~C.,  2005, \mn@doi [\apj] {10.1086/429402},
  \href {https://ui.adsabs.harvard.edu/abs/2005ApJ...625..427L} {625, 427}

\bibitem[\protect\citeauthoryear{{Leinhardt}, {Dobinson}, {Carter}  \&
  {Lines}}{{Leinhardt} et~al.}{2015}]{2015ApJ...806...23L}
{Leinhardt} Z.~M.,  {Dobinson} J.,  {Carter} P.~J.,   {Lines} S.,  2015,
  \mn@doi [\apj] {10.1088/0004-637X/806/1/23}, \href
  {https://ui.adsabs.harvard.edu/\#abs/2015ApJ...806...23L} {806, 23}

\bibitem[\protect\citeauthoryear{{Lissauer}}{{Lissauer}}{1993}]{1993ARA&A..31..129L}
{Lissauer} J.~J.,  1993, \mn@doi [\araa] {10.1146/annurev.aa.31.090193.001021},
  \href {https://ui.adsabs.harvard.edu/abs/1993ARA&A..31..129L} {31, 129}

\bibitem[\protect\citeauthoryear{{Malhotra}}{{Malhotra}}{1994}]{1994PhyD...77..289M}
{Malhotra} R.,  1994, \mn@doi [Physica D Nonlinear Phenomena]
  {10.1016/0167-2789(94)90141-4}, \href
  {https://ui.adsabs.harvard.edu/abs/1994PhyD...77..289M} {77, 289}

\bibitem[\protect\citeauthoryear{{McMullin}, {Waters}, {Schiebel}, {Young}  \&
  {Golap}}{{McMullin} et~al.}{2007}]{2007ASPC..376..127M}
{McMullin} J.~P.,  {Waters} B.,  {Schiebel} D.,  {Young} W.,   {Golap} K.,
  2007, in {Shaw} R.~A.,  {Hill} F.,   {Bell} D.~J.,  eds,  Astronomical
  Society of the Pacific Conference Series Vol. 376, Astronomical Data Analysis
  Software and Systems XVI. p.~127

\bibitem[\protect\citeauthoryear{{Menon}, {Wesolowski}  \& {Zheng}}{{Menon}
  et~al.}{2015}]{2015AphCom..2..1}
{Menon} H.,  {Wesolowski} L.,   {Zheng} G. e.~a.,  2015, Computational
  Astrophysics and Cosmology, \href
  {http://dx.doi.org/10.1186/s40668-015-0007-9} {2, 1}

\bibitem[\protect\citeauthoryear{{Murray} \& {Dermott}}{{Murray} \&
  {Dermott}}{1999}]{1999ssd..book.....M}
{Murray} C.~D.,  {Dermott} S.~F.,  1999, {Solar system dynamics}

\bibitem[\protect\citeauthoryear{{Nesvold} \& {Kuchner}}{{Nesvold} \&
  {Kuchner}}{2015}]{2015ApJ...798...83N}
{Nesvold} E.~R.,  {Kuchner} M.~J.,  2015, \mn@doi [\apj]
  {10.1088/0004-637X/798/2/83}, \href
  {https://ui.adsabs.harvard.edu/abs/2015ApJ...798...83N} {798, 83}

\bibitem[\protect\citeauthoryear{{Peale}}{{Peale}}{1976}]{1976ARA&A..14..215P}
{Peale} S.~J.,  1976, \mn@doi [\araa] {10.1146/annurev.aa.14.090176.001243},
  \href {https://ui.adsabs.harvard.edu/abs/1976ARA&A..14..215P} {14, 215}

\bibitem[\protect\citeauthoryear{{Pearce} \& {Wyatt}}{{Pearce} \&
  {Wyatt}}{2014}]{2014MNRAS.443.2541P}
{Pearce} T.~D.,  {Wyatt} M.~C.,  2014, \mn@doi [\mnras]
  {10.1093/mnras/stu1302}, \href
  {https://ui.adsabs.harvard.edu/abs/2014MNRAS.443.2541P} {443, 2541}

\bibitem[\protect\citeauthoryear{{Pearce}, {Wyatt}  \& {Kennedy}}{{Pearce}
  et~al.}{2015}]{2015MNRAS.448.3679P}
{Pearce} T.~D.,  {Wyatt} M.~C.,   {Kennedy} G.~M.,  2015, \mn@doi [\mnras]
  {10.1093/mnras/stv252}, \href
  {https://ui.adsabs.harvard.edu/abs/2015MNRAS.448.3679P} {448, 3679}

\bibitem[\protect\citeauthoryear{Pedregosa et~al.,}{Pedregosa
  et~al.}{2011}]{scikit-learn}
Pedregosa F.,  et~al., 2011, Journal of Machine Learning Research, 12, 2825

\bibitem[\protect\citeauthoryear{{P{\'e}rez} et~al.,}{{P{\'e}rez}
  et~al.}{2016}]{2016Sci...353.1519P}
{P{\'e}rez} L.~M.,  et~al., 2016, \mn@doi [Science] {10.1126/science.aaf8296},
  \href {https://ui.adsabs.harvard.edu/abs/2016Sci...353.1519P} {353, 1519}

\bibitem[\protect\citeauthoryear{{Pontzen}, {Ro{\v{s}}kar}, {Stinson}  \&
  {Woods}}{{Pontzen} et~al.}{2013}]{2013ascl.soft05002P}
{Pontzen} A.,  {Ro{\v{s}}kar} R.,  {Stinson} G.,   {Woods} R.,  2013, {pynbody:
  N-Body/SPH analysis for python} (\mn@eprint {ascl} {1305.002})

\bibitem[\protect\citeauthoryear{{Ricci}, {Isella}, {Liu}  \& {Li}}{{Ricci}
  et~al.}{2018}]{2018ASPC..517..147R}
{Ricci} L.,  {Isella} A.,  {Liu} S.,   {Li} H.,  2018, in {Murphy} E.,  ed.,
  Astronomical Society of the Pacific Conference Series Vol. 517, Science with
  a Next Generation Very Large Array. p.~147

\bibitem[\protect\citeauthoryear{{Richardson}}{{Richardson}}{1994}]{1994MNRAS.269..493R}
{Richardson} D.~C.,  1994, \mn@doi [\mnras] {10.1093/mnras/269.2.493}, \href
  {https://ui.adsabs.harvard.edu/abs/1994MNRAS.269..493R} {269, 493}

\bibitem[\protect\citeauthoryear{{Richardson}, {Quinn}, {Stadel}  \&
  {Lake}}{{Richardson} et~al.}{2000}]{2000Icar..143...45R}
{Richardson} D.~C.,  {Quinn} T.,  {Stadel} J.,   {Lake} G.,  2000, \mn@doi
  [\icarus] {10.1006/icar.1999.6243}, \href
  {https://ui.adsabs.harvard.edu/\#abs/2000Icar..143...45R} {143, 45}

\bibitem[\protect\citeauthoryear{{Safronov}}{{Safronov}}{1969}]{1969edo..book.....S}
{Safronov} V.~S.,  1969, {Evoliutsiia doplanetnogo oblaka.}

\bibitem[\protect\citeauthoryear{{Tabeshian} \& {Wiegert}}{{Tabeshian} \&
  {Wiegert}}{2016}]{2016ApJ...818..159T}
{Tabeshian} M.,  {Wiegert} P.~A.,  2016, \mn@doi [\apj]
  {10.3847/0004-637X/818/2/159}, \href
  {https://ui.adsabs.harvard.edu/\#abs/2016ApJ...818..159T} {818, 159}

\bibitem[\protect\citeauthoryear{{Tabeshian} \& {Wiegert}}{{Tabeshian} \&
  {Wiegert}}{2018}]{2018ApJ...857....3T}
{Tabeshian} M.,  {Wiegert} P.~A.,  2018, \mn@doi [\apj]
  {10.3847/1538-4357/aab668}, \href
  {https://ui.adsabs.harvard.edu/\#abs/2018ApJ...857....3T} {857, 3}

\bibitem[\protect\citeauthoryear{{Wallace} \& {Quinn}}{{Wallace} \&
  {Quinn}}{2019}]{2019MNRAS.489.2159W}
{Wallace} S.~C.,  {Quinn} T.~R.,  2019, \mn@doi [\mnras]
  {10.1093/mnras/stz2284}, \href
  {https://ui.adsabs.harvard.edu/abs/2019MNRAS.489.2159W} {489, 2159}

\bibitem[\protect\citeauthoryear{{Weidenschilling} \&
  {Davis}}{{Weidenschilling} \& {Davis}}{1985}]{1985Icar...62...16W}
{Weidenschilling} S.~J.,  {Davis} D.~R.,  1985, \mn@doi [\icarus]
  {10.1016/0019-1035(85)90169-1}, \href
  {https://ui.adsabs.harvard.edu/abs/1985Icar...62...16W} {62, 16}

\bibitem[\protect\citeauthoryear{{W}es {M}c{K}inney}{{W}es
  {M}c{K}inney}{2010}]{mckinney-proc-scipy-2010}
{W}es {M}c{K}inney 2010, in {S}t\'efan van~der {W}alt {J}arrod {M}illman eds,
  {P}roceedings of the 9th {P}ython in {S}cience {C}onference. pp 56 -- 61,
  \mn@doi{10.25080/Majora-92bf1922-00a}

\bibitem[\protect\citeauthoryear{{Wyatt}}{{Wyatt}}{2008}]{2008ARA&A..46..339W}
{Wyatt} M.~C.,  2008, \mn@doi [Annual Review of Astronomy and Astrophysics]
  {10.1146/annurev.astro.45.051806.110525}, \href
  {https://ui.adsabs.harvard.edu/abs/2008ARA&A..46..339W} {46, 339}

\bibitem[\protect\citeauthoryear{{Wyatt}, {Dermott}, {Telesco}, {Fisher},
  {Grogan}, {Holmes}  \& {Pi{\~n}a}}{{Wyatt}
  et~al.}{1999}]{1999ApJ...527..918W}
{Wyatt} M.~C.,  {Dermott} S.~F.,  {Telesco} C.~M.,  {Fisher} R.~S.,  {Grogan}
  K.,  {Holmes} E.~K.,   {Pi{\~n}a} R.~K.,  1999, \mn@doi [\apj]
  {10.1086/308093}, \href
  {https://ui.adsabs.harvard.edu/abs/1999ApJ...527..918W} {527, 918}

\bibitem[\protect\citeauthoryear{{van der Walt}, {Colbert}  \&
  {Varoquaux}}{{van der Walt} et~al.}{2011}]{2011CSE....13b..22V}
{van der Walt} S.,  {Colbert} S.~C.,   {Varoquaux} G.,  2011, \mn@doi
  [Computing in Science and Engineering] {10.1109/MCSE.2011.37}, \href
  {https://ui.adsabs.harvard.edu/abs/2011CSE....13b..22V} {13, 22}

\makeatother
\end{thebibliography}

\appendix
\section{Timescale for Secular Forcing}\label{sec:sec_forcing_timescale}

The timescale for secular forcing is given by the period at which the free eccentricity of a body cycles in the complex plane. The time-dependent imaginary and real components of 
the eccentricity of a planetesimal are given by \citet{1999ssd..book.....M} (see ch. 7, pg. 282, eq. 7.51)

\begin{eqnarray}\label{eq:kandh}
	h = e_{1} sin (g_{1} t + \beta_{1}) + e_{2} sin (g_{2} t + \beta_{2}) \\ \nonumber
	k = e_{1} cos (g_{1} t + \beta_{1}) + e_{2} cos (g_{2} t + \beta_{2}),
\end{eqnarray}

\noindent where  ($e_{1}$, $e_{2}$) and ($g_{1}$ , $g_{2}$) are the corresponding eigenvector components and eigenvalues of a matrix, which describes the secular interaction 
between two bodies, given by \citet{1999ssd..book.....M} (see ch. 7, pg. 276, eq. 7.9 and 7.10, with $\alpha_{12} = \alpha$ and $\bar{\alpha_{12}} = 1$)

\begin{eqnarray}\label{eq:pert_matrix}
	A_{jj} = n_{j} \frac{1}{4} \frac{m_{k}}{m_{c} + m_{j}} \alpha b_{3/2}^{(1)} (\alpha) \\ \nonumber
	A_{jk} = -n_{j} \frac{1}{4} \frac{m_{k}}{m_{c} + m_{j}} \alpha b_{3/2}^{(2)} (\alpha).
\end{eqnarray}

\noindent Here, $n$ is the mean motion of a body and $m$ is its mass. In the case of a massless test particle being perturbed by a giant planet, $m_{1} = m_{g}$ and $m_{2} = 0$. 
The frequency of the secular forcing is simply the difference of the two eigenvalue frequencies $g_{1} - g_{2}$.

\section{Libration in Mean Motion Resonance}\label{sec:libration}

For second order and higher MMRs, the maximum libration width is given by \citet{1999ssd..book.....M} (ch. 8, pg. 338, eq. 8.58) as

\begin{equation}\label{eq:res_so}
	\frac{\delta a}{a} = \pm \left( \frac{16}{3} \frac{\left| C_{r} \right|}{n} e^{\left| q \right|} \right)^{1/2},
\end{equation}

\noindent where $n$, $a$ and $e$ are the mean motion, semimajor axis and eccentricity of the unperturbed body. $C_{r}$ is a constant defined by 
$m'/m_{c} n \alpha f_{d}(\alpha)$, where $\alpha$ = $a/a'$ and $f_{d}$ is the resonant part of the disturbing function. For an interior second-order 
resonance,

\begin{eqnarray}\label{eq:fd_so}
	f_{d} (\alpha) &=& \frac{1}{8} \left[ -5(p+q) + 4(p+q)^{2} - 2 \alpha D \right. \\ \nonumber
	                      & + & \left. 4(p+q) \alpha D + \alpha^{2} D^{2} \right] b^{p+q}_{1/2},
\end{eqnarray}

\noindent (see \citet{1999ssd..book.....M} table 8.1, third row, with $j = p + q$) where $D b^{j}_{s}$ is the first derivative with respect to $\alpha$ of the Laplace coefficient defined in 
equation \ref{eq:lap}. This is given by \citet{1961mcm..book.....B} as

\begin{equation}\label{eq:lap_d}
	\frac{d b_{s}^{j}}{d \alpha} = s \left( b_{s+1}^{j-1} - 2 \alpha b_{s+1}^{j} + b_{s+1}^{j+1} \right).
\end{equation}

For a first-order resonance, the associated disturbing function terms are slightly simpler such that

\begin{equation}\label{eq:fd_fo}
	f_{d}(\alpha) = -(p+q) b_{1/2}^{p+q} - \frac{\alpha}{2} D b_{1/2}^{p+q},
\end{equation}

\noindent (see \citet{1999ssd..book.....M} table 8., first row) and there is an additional contribution to the motion of the critical angle by the pericenter precession rate such that

\begin{equation}\label{eq:res_fo}
	\frac{\delta a}{a} = \pm \left(\frac{16}{3} \frac{\left| C_{r} \right|}{n} e \right)^{1/2} \left(  1 + \frac{1}{27 p^2 e^3} \frac{\left| C_{r} \right|}{n} 
	\right)^{1/2} - \frac{2}{9 p e}  \frac{\left| C_{r} \right|}{n}
\end{equation}

\noindent (see \citet{1999ssd..book.....M} ch. 8, pg. 340, eq. 8.76, with $j_{2} = p$).

In mean-motion resonance, the orbital elements of a perturbed body, along with the critical angle described by equation \ref{eq:phi_crit}, will exhibit oscillations. In the circular, 
restricted three-body case, these small amplitude oscillations exhibit a characteristic frequency, given by (see \citet{1999ssd..book.....M} ch. 8, pg. 332, eq. 8.47, with $j_{2} =-p$ and 
$j_{4} = -q$)

\begin{equation}\label{eq:lib_time}
	\omega_{0} = 3 p^{2} C_{r} n e^{\left| -q \right|}.
\end{equation}

\section{Nonaxisymmetric Structure in Boley 2017}\label{sec:boley_plot}

\begin{figure*}
    \includegraphics[width=\textwidth]{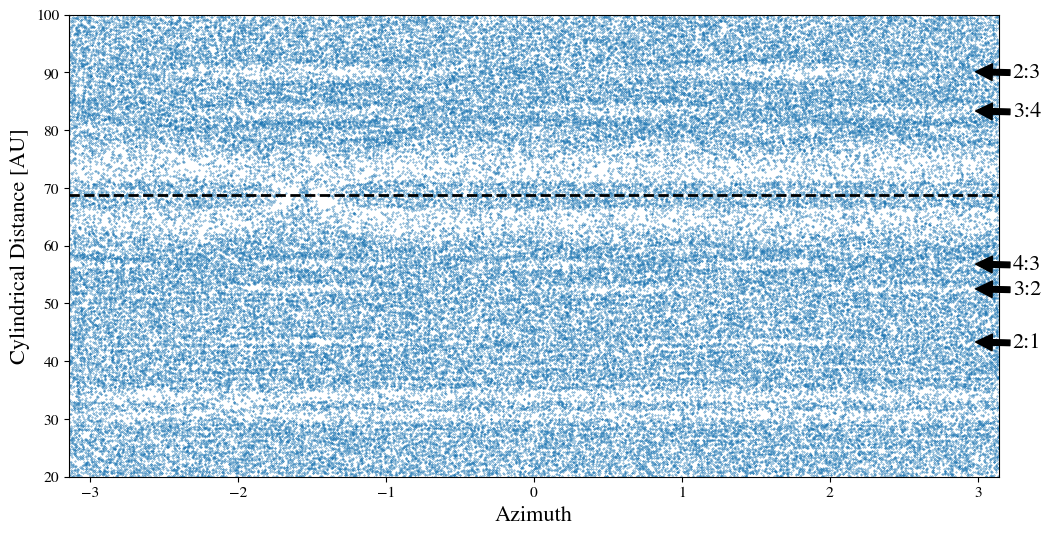}
    \caption{Positions of planetesimals at the end of the simulation presented in \citet{2017ApJ...850..103B} in polar coordinates. The dashed line marks the location of a 50 $M_{\earth}$ 
    planet embedded in the disc. First order MMRs with the planet are indicated with arrows. A second planet is also present at 32.3 au, although it is not massive enough to induce any 
    significant structure in the disc near associated resonances.\label{fig:boley_rtheta}}
\end{figure*}

As mentioned in section \ref{sec:vary_ecc}, the nonaxisymmetric gap structure seen in figure \ref{fig:xy} is present in a number of other studies,
including \citet{2000Icar..143...45R} and \citet{2016ApJ...818..159T}. One would expect the same dynamics to be at work in the simulation of 
\citet{2017ApJ...850..103B}, although the structure is not obviously visible in figure 3 of their work.

Here, we present a closer look at the positions of the planetesimals at the end of the \citet{2017ApJ...850..103B} simulation. In figure 
\ref{fig:boley_rtheta}, we show the planetesimal positions in polar, rather than Cartesian coordinates as originally presented in the paper. Two planets 
on circular orbits are present here. The first has mass of 17 $M_{\earth}$ and lies at 32.3 au. The second planet has a mass of 50 $M_{\earth}$ 
and lies at 68.8 au. A horseshoe shaped collection of planetesimals is visible to be corotating with each planet. Nonaxisymmetric gaps are also
visible at the locations of first order MMRs with the more massive planet, which are indicated with arrows.

Similar to figure \ref{fig:coll_polar_e}, a gap is present at the same orbital phase as the planet at the 2:1 MMR, along with a second gap on
the opposite side of the disc. Gaps near the 4:3 and 3:2 MMRs are also present in phase with the planet, although the number and spacing between 
additional gaps appears to be different for each resonance. In addition, the gap structure associated with the exterior MMRs appears markedly
different than that of each corresponding interior MMR, which suggests that the structure is related to the symmetry of the path that a planetesimal
in a frame corotating with the planet takes.

\bsp	
\label{lastpage}
\end{document}